\documentclass[aps,prb,twocolumn,showpacs,superscriptaddress]{revtex4}

\usepackage{natbib} \usepackage{graphicx} \usepackage{amsmath}
\usepackage{bm} \usepackage[colorlinks]{hyperref} \usepackage{color}

\newcommand{\EqLabel}[1]{\label{#1}} \newcommand{\mb}[1]{\mathbf{#1}}

\newcommand{\kh}{\frac{k}{2}} \newcommand{\st}[1]{{\scriptstyle #1}}
\newcommand{\crdu}[1]{c_{#1,\uparrow}^{\dagger}}
\newcommand{\crdd}[1]{c_{#1,\downarrow}^{\dagger}}

\newcommand{\ckdu}[1]{c_{#1,\uparrow}^{\dagger}}
\newcommand{\ckdd}[1]{c_{#1,\downarrow}^{\dagger}}

\newcommand{\ckd}[1]{c_{#1,\downarrow}}

\begin{document}
 
\title{The role of the lattice structure in determining the
  magnon-mediated interactions between charge carriers doped into a
  magnetically ordered background}

\author{Mirko M\" oller} 
\affiliation{Department of Physics and Astronomy, University of British
  Columbia, Vancouver, BC, Canada, V6T 1Z1} 
\affiliation{Department of Physics, Freie Universit\"at Berlin,
  Arnimallee 14, 14195 Berlin, Germany} 

\author{George A. Sawatzky}
\affiliation{Department of Physics and Astronomy, University of British
  Columbia, Vancouver, BC, Canada, V6T 1Z1} 
\affiliation{Quantum Matter Institute, University of British
  Columbia, Vancouver, BC, Canada, V6T 1Z4} 
 
\author{Mona Berciu}
\affiliation{Department of Physics and Astronomy, University of British
  Columbia, Vancouver, BC, Canada, V6T 1Z1} 
\affiliation{Quantum Matter Institute, University of British
  Columbia, Vancouver, BC, Canada, V6T 1Z4} 

\date{\today}
 
\begin{abstract} 
We use two recently proposed methods to calculate exactly the spectrum
of two spin-${1\over 2}$ charge carriers moving in a ferromagnetic
background, at zero temperature, for three types of models. By
comparing the low-energy states in both the one-carrier and the
two-carrier sectors, we analyze whether complex models with multiple
sublattices can be accurately described by simpler Hamiltonians, such
as one-band models. We find that while this is possible in the
one-particle sector, the magnon-mediated interactions which are key to
properly describe the two-carrier states of the complex model are not
reproduced by the simpler models. We argue that this is true not just
for ferromagnetic, but also for antiferromagnetic backgrounds. Our
results question the ability of simple one-band models to accurately describe
the low-energy physics of cuprate layers.
\end{abstract}


\maketitle

\section{Introduction}

Many materials under current study, such as cuprates, manganites,
pnictides, irridates, etc., have complicated structures with several
types of atoms in the basis. Including orbitals for all these atoms in
a model Hamiltonian would make it impossibly difficult to solve,
besides introducing unreasonably many parameters.

Thus, the first major challenge in studying such materials is to
understand what is the minimal model Hamiltonian that properly
captures their low-energy properties. (The second major challenge, of
course, is to figure how to solve it). Some steps in this process are
fairly straightforward. For instance, many of these materials have
layered structures, and there are many indications that the
interesting physics is hosted by certain layers which are common to
all compounds in the family. It is therefore an easy decision to start
by modeling one such layer -- for example, a CuO$_2$ layer for
cuprates.

Even though this is a significant simplification, the proper minimal
model to describe the low-energy properties of one layer is still
not obvious. To continue with the CuO$_2$ example, one could write a
model that includes the $3d_{x^2-y^2}$ orbitals for Cu, which are
known to be half-filled in the undoped parent compound, plus the
appropriate $2p_{x/y}$ O orbitals. These are filled in the undoped
parent compounds, but the cuprates are charge transfer
insulators\cite{ZSA} and these are the orbitals expected to host the
holes introduced by doping. This approach leads to the three-band model of
Emery,\cite{E3b} although for later convenience we prefer to label it
here as a two-sublattice model: one for the Cu sites, effectively
hosting spins-${1\over 2}$ if double occupancy is forbidden, and one
for the O sites, hosting the  holes  (the charge carriers).

A simpler option is a one-band Hubbard model which describes
effective states located on one lattice (at the Cu sites, for CuO$_2$
layers), or its even simpler counterpart, the $t$-$J$ model, which
additionally discards all doubly occupied states and describes carriers in a
spin background. Even these simplest possible interacting
Hamiltonians, which depend on only one dimensionless parameter ($U/t$
or $J/t$, respectively) are still  far from completely understood
in the strongly correlated limit, despite a tremendous amount of
effort.\cite{hub}

It is the difficulty to solve such correlated Hamiltonians that explains the
drive to simplify the models that we study to the utmost
possible. However, it is reasonable to expect that as more and more
details are discarded, there is a point beyond which the model no
longer contains the physics that we aim to study. Are the
Hubbard or $t$-$J$ models already past this point, as far as cuprates
are concerned, or do they still contain the low-energy physics of the
more complex Emery model?

An affirmative answer to the latter question was given by Zhang and
Rice,\cite{ZR} who also clarified the nature of the states appearing
in these one-band models, namely the Zhang-Rice singlets (ZRS). We
review their arguments in some detail below; for now it suffices to
say that the ZRS is a composite object involving a doping hole
occupying a certain linear combination of O orbitals, and which is
locked in a singlet with a hole (spin) at a Cu site.

Zhang and Rice's work on the equivalence between the low-energy
properties of the Emery and $t$-$J$ models is based on arguments
limited to the one-carrier (one doping hole) sector. Its validity is
still not fully settled because of the lack of exact results for these
Hamiltonians. Accurate analytical approximations are not available,
while conclusions based on numerical work are hampered by the
restriction to rather small clusters. Recent results for clusters with
32 CuO$_2$ unit cells suggest that at least in parts of the Brillouin
zone (BZ), the quantum numbers and symmetries of the lowest eigenstate
of the Emery model are different from those of its $t$-$J$
counterpart, although the energy dispersion is rather
similar.\cite{Bayo1}

While efforts to understand if the ZRS offers a good description of the
quasiparticle are on-going, it is important to note that even if this
is proven true, it would still not settle the question whether
one-band models give a proper description of the low-energy properties
of weakly doped CuO$_2$ layers. This is because a model Hamiltonian
must describe properly not only 
individual quasiparticles, but also {\em their interactions}. This is
especially important in cuprates where high-temperature
superconductivity appears upon doping. Much of the community believes
the pairing to be mediated through magnon exchange,\cite{magnons}
although phonon-mediated pairing\cite{phonons} and even
combined mechanisms\cite{italians} are also favored. Clearly, the
model Hamiltonian must  describe properly the magnon-mediated
interactions before one can decide if they are sufficient to
facilitate pairing and superconductivity at these rather high
temperatures, or whether coupling to additional bosons, {\em e.g.}
phonons, needs to be considered.

To settle this issue, one must compare the low-energy properties of
the two models in the two-carrier sector; the single-carrier sector
has no information about interactions. For the Emery and $t$-$J$ models,
such a comparison is even more difficult, since finite size effects
become more important as the number of holes increases. Recent efforts
in this direction were rather inconclusive.\cite{Bayo2}

Although the discussion so far was focused on cuprates, similar
questions appear for other materials. Is it reasonable to
ignore\cite{man} the ligand O and use models for manganites which only
involve Mn $3d$ orbitals? Do the As ions play any role\cite{pnic} in
the low-energy physics of pnictides? These and many other similar
questions have, as a common denominator, the issue of whether one has
to use complex models with different ions on different sublattices to
properly describe such materials, or whether it suffices to use
one-lattice models, possibly based on some composite states which
effectively account for the role played by other ions, such as the
ligands.

\begin{figure}[b]
\centering \includegraphics[angle=0, width=0.4\columnwidth]{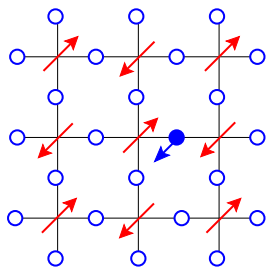}
  \caption{(Color online) Sketch of a CuO$_2$ plane. If double occupancy is
  forbidden, each Cu site has a hole in a $3d_{x^2-y^2}$ orbital,
  behaving like a spin-${1\over 2}$ (red arrows). Upon doping, the O
  $2p_{x/y}$ orbitals (empty blue dots) accommodate the doping holes
  (filled blue dots, with arrows indicating the spins).}
  \label{fig1}
\end{figure}

Here we focus on one class of such questions, for materials where the
parent compound has long range magnetic order and is a charge-transfer
insulator. (A short version of this work was published in
Ref. \onlinecite{old}). Our conclusions are based on results for the one- and
two-carrier sectors for a background that has ferromagnetic (FM)
order. This is much simpler than an antiferromagnetic (AFM) background
and allows us to find exact solutions, so that we are certain that
differences between results are due to the models themselves. We find
that the simpler one-lattice models {\em  do not} properly describe
magnon-mediated 
interactions between carriers, and identify the simple reasons for
this. Moreover, we argue that these conclusions are relevant for AFM
backgrounds as well. We believe that our results offer strong
arguments that the one-band Hubbard and  $t-J$ models  are
inappropriate to describe cuprate layers.

The work is organized as follows. In Section II we briefly review the
ZRS arguments. In Section III we introduce our models, which mirror
the steps in the ZRS derivation, but for a FM background. Section IV
summarizes the formalism (details are presented in various
Appendixes). The results are in Section V, and  Section VI
contains a detailed discussion and our conclusions.

\section{Brief review of the ZRS}

The Emery model for a CuO$_2$ plane, sketched in Fig. \ref{fig1},
includes explicitly both the Cu and the O sublattices, with their
$3d_{x^2-y^2}$ and $2p_{x/y}$ orbitals, respectively.  The goal is to
try to reduce it to a one-lattice model, built on effective states
located at the Cu sites.

Since the doping hole occupies O $2p$ orbitals, the first step to
remove the O sublattice is to make linear combinations of the four O
orbitals neighboring one Cu, which are thus centered at the Cu
sites $i$. These are described by the $P_{i,\sigma}^{(S,A)}$ operators in
Eq. (5) of Ref. \onlinecite{E3b}. Because
of phase coherence, the ``S'' combination where the O
orbitals have ${x^2-y^2}$ symmetry is the low-energy
state, and from now on we restrict the discussion to it.

The $P_{i,\sigma}^{(S)}$ states are not orthogonal since neighbor Cu
share an O ligand, [Eq. (6) of Ref. \onlinecite{E3b}]. To fix this,
Zhang and Rice first define their Fourier transforms $P_{\mb{k},
  \sigma} = {1\over \sqrt{N_S}} \beta_{\mb{k}} \sum_{i}^{} e^{ -i
  \mb{k} \mb{R}_i} P^{(S)}_{i,\sigma}$, where the normalization factor
$\beta_{\mb k} = 1/\sqrt{1- {1\over 2} ( \cos k_x + \cos k_y)}$ is due
to the non-orthogonality of the $P^{(S)}_{i,\sigma}$ [Eqs. (8) and (9)
  of Ref. \onlinecite{E3b}]. It is problematic that $\beta_{\mb k}$
diverges near the center of the BZ, but ignoring this fact one can
Fourier transform back to real space to obtain the operators
$\phi_{i,\sigma} = {1\over \sqrt{N_S}} \sum_{i}^{} e^{ i \mb{k}
  \mb{R}_i} P_{{\mb k}\sigma}$. These are now properly  normalized and
describe ``effective Cu orbitals'', although they are in fact
rather complicated linear combinations of O orbitals. The next step is
to note that exchange favors locking a doping hole occupying such an orbital
in a singlet with the hole (spin) located on the central Cu. This
results in the ZRS 
described by the operators $\psi_i = {1\over \sqrt{2}}
\left(\phi_{i\uparrow}d_{i\downarrow} - \phi_{i\downarrow}
d_{i\uparrow} \right)$, see Eq. (10) of Ref. \onlinecite{E3b}. Zhang
and Rice then argue that if double occupancy is forbidden, a $t$-$J$
Hamiltonian based on these ZRS captures  the
low-energy states of the Emery model.

\section{Models} \label{sec:Models}

\begin{figure}[t]
\centering \includegraphics[angle=0, width=0.6\columnwidth]{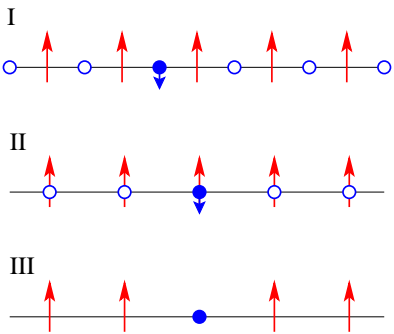}
  \caption{(Clor online) Sketches of the three models. Models I and
  model II have two bands. One (empty circles) hosts charge carriers
  introduced by doping (filled circle with arrow for the carrier spin)
  and the other one hosts lattice spins (red arrows). In model I the
  carriers propagate on a different sublattice than the one hosting
  the spins, while in model II, they are on the same lattice. Model
  III has only one band which hosts both spins (arrows) and ZRS-like
  objects (filled circle).}
  \label{fig2}
\end{figure}

In an effort to mirror these steps in going from the Emery to the
 $t$-$J$ model, we study three cases sketched in
 Fig. \ref{fig2}. For the sake of simplicity we restrict ourselves to
 one dimension (1D).  Generalizations to higher dimensions are
 straightforward, but as explained below, they do not lead to
 qualitative changes.

Case I is the ``parent'' two-sublattice, two-band model. One
sublattice hosts the lattice spins, of quantum number $S$, while the
other hosts the charge carriers. This is the 1D counterpart of the
Emery model for the 2D CuO$_2$ layers if double occupancy is forbidden
(apart from the FM order, of course).  Model II is a one-lattice,
two-band model. One should roughly think of these on-site carrier orbitals as
the analogs of the $\phi_{i,\sigma}$ in the ZRS
derivation. Finally, model III is an even simpler one-band model, the
equivalent of the $t$-$J$ model obtained after locking doping holes in
singlets with lattice spins.  These parallels will be made more clear
in the Results section.

In all these models lattice spins interact ferromagnetically with
their nearest neighbors:
\begin{align}
 \mathcal{H}_{S}=-J\sum_i (\vec{S}_i \vec{S}_{i+1}-S^2),
\label{e1}
\end{align}
where $\vec{S}_i$ is the operator for the spin at
 site $R_i=i a$, and $a=1$ is the lattice constant. Its ground-state
 is $| {\rm FM}\rangle = |+S,+S,\dots
 +S\rangle$ and has zero energy. For reasons detailed below, the
 excited states of interest will be the one-magnon states $S_q^-| {\rm FM}
 \rangle=\frac{1}{\sqrt{2SN}}\sum_i e^{iqR_i} S_i^-| {\rm FM} \rangle$ with
 energy $\Omega_q=4JS \sin^2 \left (\frac{q}{2} \right)$. The raising
 and lowering operators $S_i^\pm=S_i^x\pm i S_i^y$ correspond to
 creation $(-)$ and annihilation $(+)$ operators for magnons,
 respectively.  If we rewrite
 $\mathcal{H}_{S}=\mathcal{H}_{S}^z+\mathcal{H}_{S}^{x,y}$, where
 $\mathcal{H}_{S}^z=-J\sum_i (S_i^z S_{i+1}^z-S^2)$ and
 $\mathcal{H}_{S}^{x,y}=-\frac{J}{2} \sum_{i} \left (S_{i}^{+}
 S_{i+1}^{-} + S_{i}^{-} S_{i+1}^{+} \right )$, it is clear that
 $\mathcal{H}_{S}^{x,y}$ describes magnon hopping.

In models I and II, the second band hosts spin-$\frac{1}{2}$ charge
carriers (electrons or holes) introduced by doping. For model II these
carriers propagate on the same lattice as the spins. For model I, they
move on a different sublattice which is interlaced with the
spin sublattice (see Fig. \ref{fig2}). For both models the carriers
are described by a Hubbard Hamiltonian
$\mathcal{H}_{C}=\hat{T}+\hat{U}$, where:
\begin{align}
 &\hat{T}=-t\sum_{i,\sigma} ( c_{i+\delta+1, \sigma}^\dagger
 c_{i+\delta,\sigma}+\text{H.c.}), \\ &\hat{U}= U \sum_{i}
 n_{i+\delta, \uparrow} n_{i+\delta, \downarrow},
\end{align}
and $\delta=\frac{1}{2}$ for model I, $\delta=0$ for model II. Here,
$c_{i+\delta,\sigma}^\dagger$ is the creation operator for a carrier
with spin $\sigma$ at site $i+\delta$, and $n_{i+\delta,
\sigma}=c_{i+\delta,\sigma}^\dagger c_{i+\delta,\sigma}$ is the
occupation number operator. The hopping $\hat{T}$ is diagonalized by the
states $c_{k, \sigma}^{\dagger}=\frac{1}{\sqrt{N}} \sum_{i}e^{i k
R_{i+\delta}} c_{i+\delta,\sigma}^{\dagger}$, where $N$ is the number
of unit cells. The resulting free carrier 
dispersion is $\epsilon(k)=-2t\cos (k)$, where $k\in (-\pi, \pi]$ is
in the BZ.

The exchange between the carriers' spins and the lattice spins takes
the simplest form:
$$ \mathcal{H}_{ex}^{(I)}=J_0 \sum_{i} \vec{s}_{i+\frac{1}{2}} (
\vec{S}_{i}+\vec{S}_{i+1})
$$ for model I, while for model II:
$$\mathcal{H}_{ex}^{(II)}=J_0 \sum_{i} \vec{s}_{i}\vec{S}_{i}.
$$ Here, $\vec{s}_{i+\delta}=\sum_{\alpha, \beta} c_{i+\delta,
\alpha}^\dagger \frac{\vec{\sigma}_{\alpha, \beta}}{2} c_{i+\delta,
\beta}$ is the spin operator for carriers at site $i+\delta$, and
$\vec{\sigma}$ are Pauli matrices. In the following we choose $J_0>0$
to be AFM, as appropriate for a CuO$_2$-like situation. This also has
the advantage that it favors an infinitely lived low-energy
quasiparticle, as detailed below. However, our solution works for FM 
exchange, $J_0<0$, as well. If we write
$\mathcal{H}_{ex}^{(I,II)}=\mathcal{H}_{ex;z}^{(I,II)}
+\mathcal{H}_{ex;x,y}^{(I,II)}$, with, e.g.,
$\mathcal{H}_{ex;z}^{(II)}=J_{0} \sum_{i} s_{i}^{z} S_{i}^{z}$ and
$\mathcal{H}_{ex;x,y}^{(II)}=\frac{J_{0}}{2} \sum_{i} \left [ s_i^+
S_{i}^{-} + s_i^- S_{i}^{+} \right]$, etc., the latter term describes
spin-flip processes in which a charge carrier either absorbs or emits
a magnon.

While in this work we take the total Hamiltonians for cases I and II
to be
$\mathcal{H}_{I,II}=\mathcal{H}_{C}+\mathcal{H}_{S}+\mathcal{H}_{ex}^{(I,II)}$,
it is important to note that if the no double-occupancy restriction is
imposed properly, there are additional terms arising in the
two-sublattice analog of the Emery model.\cite{Bayo1, Bayo2,others}
For example, 
exchange between lattice spins is blocked on bonds where carriers
reside. Since this is of order $J$, which will be chosen as the
smallest energy scale in the problem, its effects may not be that
important. More substantial is the appearance of a ``spin-swap'' term,
which describes carrier hopping while its spin is exchanged with that
of the lattice spin past which it hops. This term has an energy scale
comparable to $J_0$, as expected since they both come from
second-order processes in the $t_{pd}$ hopping.\cite{Bayo2,others} Since
$J_0$ is a large energy scale, the effect of ignoring this term may be
substantial. Such terms can be easily handled by our exact
solution,\cite{FM} but we do not include them here so as to keep
models I and II as similar as possible. This allows us to conclude
that qualitative differences between their results are due to the different
lattice structures, whereas if such additional terms were present in one
but not the other model, this would no longer be clear. Note that
these additional terms were ignored by Zhang and Rice as well.

The one-band model III is described by a $t$-$J$ model, where the
spin-spin exchange ${\cal H}_S$ of Eq. (\ref{e1}) is supplemented by
``hole'' hopping (where the ``hole'' is a charged singlet, not a
regular fermionic charge carrier). Because all lattice spins are up,
this is completely equivalent with a Hubbard Hamiltonian; double
occupancy is automatically prevented here by Pauli's principle.

Models I and II have three energy scales: $t$, $J$ and $J_0$. We will
assume throughout that $J\ll t, J_0$. This is physically reasonable
since this super-exchange arises from 4th order processes in the $t_{pd}$
hopping, whereas $J_0$ and $t$ (which is similar to the spin-swap term
but without exchange of the spins) are second order in $t_{pd}$. Of
course, $t$ could also include a direct $t_{pp}$ contribution. Which of the
latter two is larger depends on the interplay between the Hubbard
repulsion on each of the two sublattices ($U_d$ and $U_p$, in cuprate
language) and the charge transfer energy ($\Delta$). In model III, the
$J_0$ energy scale has been integrated out. It stabilizes the
low-energy nature of the ZRS-like object but it does not influence its
dynamics, hence only the $t$ and $J$ energy scales remain.

Finally, note that all these Hamiltonians are invariant to
translations and commute with the $z$-component of the total spin
$S_{\text{tot}}^z=\sum_i (S_i^z+s_{i+\delta}^z)$. This has important
consequences for the structure of their eigenstates, as described
next.

\section{Formalism} \label{sec:Formalism}

Before we can understand what interactions arise between
quasiparticles, we need to first understand the quasiparticles
themselves. This is achieved in the single carrier sector. We review
the solution for this first, and then move on to the two-carrier sector.

\subsection{Single charge carrier sector}

Since the FM background breaks spin rotational symmetry, we need to treat
separately the cases with the charge carrier injected with spin up
and spin down.

 The former case is solved trivially at $T=0$. Due to the restriction
to the $S_{\text{tot}}^z=NS+\frac{1}{2}$ subspace, no spin-flips are
possible. The state $c^\dagger_{k,\uparrow}| {\rm FM}\rangle$ is therefore an
eigenstate of ${\cal H}_{I,II}$, with energy
$E_{k,\uparrow}=\epsilon(k)+ \gamma J_0 S$, with $\gamma=1$ for model
I and $\gamma=\frac{1}{2}$ for model II. Model III cannot
differentiate between an undoped system and a system doped with
spin-up carriers, and is therefore unable to describe the physics of
this case.

The $T=0$ solution for a single spin-down charge carrier in model II
has been known for a long time,\cite{Shastry} and has recently been
generalized to two-sublattice models like model I. \cite{FM} An exact
solution can be obtained for the Green's function:
\begin{align}
 G_{\downarrow}(k,\omega)=\langle {\rm FM} | \ckd{k} \hat{G}(\omega)
 \ckdd{k} | {\rm FM} \rangle,
\end{align}
where $\hat{G}(\omega)=[\omega-\mathcal{H}+i \eta]^{-1}$ is the
resolvent for $\mathcal{H}$, $\hbar=1$ and $\eta$ is an
infinitesimally small positive number which enforces the retardation
condition. Physically, it corresponds to the introduction of a finite
lifetime $\sim 1/ \eta$. Using a Lehmann representation:
\begin{align}
 G_{\downarrow}(k,\omega)=\sum_n\frac{ \langle {\rm FM} | \ckd{k} |
 \psi_n \rangle \langle \psi_n | \ckdd{k} | {\rm FM} \rangle}{\omega -
 \epsilon_n +i \eta},
\end{align}
we find that $G_{\downarrow}(k,\omega)$ has poles at the eigenenergies
 $\epsilon_n$ in the one-carrier, $S_{\text{tot}}^z=NS-\frac{1}{2}$
 subspace: $\mathcal{H}|\psi_n\rangle = \epsilon_n
 |\psi_n\rangle$. The weights at these poles are the overlaps between
 eigenstates $| \psi_n \rangle$ and free-particle states $\ckdd{k} |
 {\rm FM}\rangle$.

The way to calculate $G_\downarrow(k,\omega)$ is well
established.\cite{Shastry,FM} One uses the Dyson identity:
\begin{align}
 \hat{G}(\omega)=\hat{G}_0(\omega)+\hat{G}(\omega)\hat{V}
 \hat{G}_0(\omega), \label{eq:Dyson}
\end{align}
where $\hat{G}_0(\omega)$ is the resolvent for $\mathcal{H}_0=\hat{T}+
\mathcal{H}_{S}^{z}$ and $\hat{V}=\mathcal{H}-\mathcal{H}_0$. The
spin-flip part of the carrier-spin exchange,
$\mathcal{H}_{ex;x,y}^{(I,II)}$, mixes the state $\ckdd{k} | {\rm FM}
\rangle$, of energy $E_{k, \downarrow}= \epsilon(k)-\gamma J_0 S$,
with the continuum of one-magnon states $\ckdu{k-q} S_q^- | {\rm FM}
\rangle$, of energy $E_{k-q, \uparrow}+\Omega_q$. Due to the
restriction to the $NS-\frac{1}{2}$ subspace, the excitation of more
than one magnon is not allowed. The resulting two coupled equations
of motion (EOM) can be  easily solved analytically (for more details
see Refs.  \onlinecite{Shastry,FM}).

For model III, the solution in this subspace is trivial: a ``hole''
propagates freely with energy $\epsilon(k)$.

\subsection{Two charge carriers sector}

Here, there are three distinct cases: (i) both carriers are injected
with spin up, $S_{\text{tot}}^z=NS+1$; (ii) carriers are injected with
opposite spins, $S_{\text{tot}}^z=NS$; and (iii) both carriers are
injected with spin-down, $S_{\text{tot}}^z=NS-1$.

Case (i) is trivial at $T=0$, since no spin-flip processes are
possible, and therefore there is no magnon-mediated interaction. (The
Hubbard repulsion has no effect, either, since both carriers have
spin-up). The eigenstates are simply $\ckdu{k} \ckdu{k'}| {\rm FM}
\rangle$ with energy $E_{k,\uparrow}+E_{k', \uparrow}$.

The $T=0$ solution for case (ii) is discussed in detail next. We
present two methods to obtain this solution, one based on a $k$-space
formulation and one based on a real-space formulation. The former is
exemplified for model II while the latter is exemplified for model I,
however both models can and have been solved with both methods. For
model III we run into the same problem as before: since this model cannot
distinguish between lattice-spins and doped spin-up carriers, we
cannot consider this case for it.

Case (iii) turns out to be less interesting for our purposes. It is
briefly discussed at the end of Section \ref{sec:Results}.

\subsubsection{k-space solution for model II in the two-carrier,
  $S_{\text{tot}}^z=NS$ sector} 
\label{sec:k-space}

This solution is rather similar to that for
$G_\downarrow(k,\omega)$. We start by defining the following Green's
functions:
\begin{align}
 & G(k,q,q',\omega)= \langle k,q' | \hat{G}(\omega) | k,q \rangle,
\end{align}
where $| k,q\rangle =
c_{\frac{k}{2}+q,\uparrow}^{\dagger}c_{\frac{k}{2}-q,\downarrow}^{\dagger}|
{\rm FM} \rangle$ is a two-carrier state with total momentum $k$ and
$S_{\text{tot}}^z=NS$. While the total momentum $k$ is conserved, $q$
is subject to change due to on-site Hubbard scattering and to
magnon-mediated interactions. The latter describe
the emission of a magnon by the spin-down carrier and its subsequent
absorption by the spin-up carrier. As a result, the two carriers
exchange their spins, besides exchanging some momentum. 

In order to construct the EOM for $G(k,q,q',\omega)$ we split the
Hamiltonian into two parts, $\mathcal{H}=\mathcal{H}_0+\hat{V}$ where
$\mathcal{H}_0=\hat{T}+\mathcal{H}_{S}^{z}$, and repeatedly use
Dyson's identity, Eq. (\ref{eq:Dyson}). After using it
once we obtain:
\begin{align}
 & G(k,q,q',\omega)=\left
 \{\delta_{q,q'}+\frac{U}{N}\sum_QG(k,Q,q',\omega) \right . \nonumber
 \\ & \left .  +\frac{J_0}{2N}\sum_pF(k,q,q',p,\omega)\right
 \}G_0(\st{\kh+q},\st{\kh-q}, \omega). \label{eq:k-spaceG1}
\end{align}
Here $G_{0}(k,k',\omega)=[\omega -\epsilon(k)-\epsilon(k')+i
  \eta]^{-1}$ is the propagator for two non-interacting carriers, and
  $F(k,q,q',p,\omega) = \sum_ie^{ipR_i} \langle k, q'| \hat{G}(\omega)
  \ckdu{\kh+q} \ckdu{\kh-q-p} S_p^- | {\rm FM} \rangle$ is a generalized
  Green's function related to states where both carriers have spin up
  and a magnon is present with momentum $p$. Such states appear when
  the spin-down carrier flips its spin to create a magnon. This
  process is mediated by $\mathcal{H}_{ex;x,y}^{(II)}$ and leaves
  $S_{\text{tot}}^z$ unchanged.

We now use Dyson's equation again to obtain the EOM for $F(k,q,q',p,\omega)$:
\begin{widetext}
\begin{align}
 F(k,q,q',p,\omega)&=\left \{-\frac{J_0}{2N}\sum_Q \left
  [F(k,q,q',Q,\omega)-F(k,-q-p,q',Q,\omega) \right]
  +J_0S[G(k,q,q',\omega) \right. \nonumber \\
  &\left. \vphantom{\sum_Q} -G(k,-q-p,q',\omega)]\right \}
  G_0(\st{\kh+q},\st{\kh-q-p},\omega-\Omega_p-J_0S). \label{eq:k-spaceF}
\end{align}
\end{widetext}

Note that we obtain only two coupled EOM, since the initial states are
linked only to states with two spin-up carriers and one magnon.
Furthermore, we do not need to know the full $F$ to solve for $G$, only its
average $\frac{1}{N}\sum_pF(k,q,q',p,\omega)$. We therefore need to
express this average in terms of $G$ and $G_0$. This allows us to
eliminate one of these equations by  first rewriting
Eq. (\ref{eq:k-spaceG1}) as $ \frac{J_0}{2N}
\sum_QF(k,q,q',Q,\omega)=-\delta_{q,q'}+G(k,q,q',\omega)G_0^{-1}
(\st{\kh+q},\st{\kh-q},\omega)-  
\frac{U}{N}\sum_QG(k,Q,q',\omega).$ Inserting this into
Eq. (\ref{eq:k-spaceF}), taking the sum over $p$\ and then inserting
the result into Eq. (\ref{eq:k-spaceG1}) we find:
\begin{widetext}
 \begin{align}
   G(k,q,q',\omega)&=\left \{ \delta_{q,q'}
\left[1+\frac{J_0}{2}g(k,q,\omega) \right] +\frac{1}{N}\sum_Q
G(k,Q,q',\omega)\left [U+\frac{J_0}{2}
\frac{G_0(\st{\kh+q},\st{\kh+Q},\tilde{\Omega}_{q+Q})}{G_0(\st{\kh+Q},
\st{\kh-Q},\omega-J_0S)} \right] \right. \nonumber \\ &
\left. \vphantom{\sum_Q}
-\frac{J_0}{2N}G_0(\st{\kh+q},\st{\kh+q'},\tilde{\Omega}_{q+q'})\right
\}
G_0\left(\st{\kh+q},\st{\kh-q},\st{\omega+\frac{J_0}{2}\frac{g(k,q,\omega)}
{G_0(\st{\kh+q},\st{\kh-q,\omega-J_0S})}}\right)
\label{eq:k-spaceG2},
 \end{align}
\end{widetext}
where we used the shorthand  $\tilde{\Omega}_q\equiv
\omega-\Omega_q-J_0S$, and $g(k,q,\omega)={1}/{N}\sum_QG_0(\st{\kh+q},
\st{\kh+Q},\tilde{\Omega}_{q+Q})$ can be calculated numerically.  To
solve Eq. (\ref{eq:k-spaceG2}) we use the fact that for a chain with
$N$ unit cells and periodic boundary conditions, the allowed momenta
are discrete: $q_i=\frac{2 \pi}{N}i$, $i=0,\dots, N-1$.  Therefore,
for any given values of $k$ and $q'$, Eq. (\ref{eq:k-spaceG2}) is a
$N$-dimensional system of linear equations associated with $q_i$, and
we can use basic linear algebra algorithms to solve it numerically. Of
course, one could have used the finite $N$ much sooner and written
Eqs. (\ref{eq:k-spaceG1}) and (\ref{eq:k-spaceF}) as two coupled
linear systems of equations. However, by removing $F$ from
Eqs. (\ref{eq:k-spaceG1}) and (\ref{eq:k-spaceF}) to obtain
Eq. (\ref{eq:k-spaceG2}), we reduced this to a system with $N$
unknowns. Otherwise, we would have to deal with an additional
$N^2$ variables for $F$. This is possible but increases the
computation time tremendously.

As already mentioned, model I can be solved similarly (for details,
see the supplementary material of Ref. \onlinecite{old}). In both
cases, results can be easily obtained for systems with $N\sim 10^2$,
so finite-size effects can be avoided. However, note that for these
$k$-space solutions, it is important that $N$ be not too large. The
reason is that we are looking for bound states, and their overlap with
the $|k,q\rangle$ extended states vanishes like $1/\sqrt{N}$, and
therefore they become ``invisible'' in the thermodynamic limit.
This problem is avoided in a real-space formulation, as discussed next.

\subsubsection{Real-space solution for model I in the two-carrier,
  $S_{\text{tot}}^z=NS$ sector} 

This solution is exemplified here for model I. It relies on a
real-space formulation based on  methods introduced in
Ref. \onlinecite{FewP}. A short discussion of this solution for model
II is available in  the supplementary material of
Ref. \onlinecite{old}.

Here we want to calculate the Green's functions:
$$
G(n,n',k,\omega)=\langle k, n' | \hat{G}(\omega) | k,
 n\rangle$$
where the states
%%%%%%%%%%%%%%%%%%%%%%%%%%%%%% EQUATION %%%%%%%%%%%%%%%%%%%%%%%%%%%%%%
\begin{equation}
\label{eq:ElBetwSpinRealSpaceState1}
 | k, n\rangle =  \sum_{i} \frac{e^{i k
 (R_{i+\frac{1}{2}}+\frac{n}{2})}}{\sqrt{N}}
 \crdu{i+\frac{1}{2}}\crdd{i+\frac{1}{2}+n} | {\rm FM} \rangle
\end{equation}
%%%%%%%%%%%%%%%%%%%%%%%%%%%%%%%%%%%%%%%%%%%%%%%%%%%%%%%%%%%%%%%%%%%%%%
describe configurations with total momentum $k$, and where the
carriers have different spins and are located $- \infty < n < \infty $
sites apart. 

To generate their EOM, we again use Dyson's identity, but now we
choose
$\mathcal{H}_0=\hat{U}+\mathcal{H}_{S}^{z}+\mathcal{H}_{ex;z}^{(I)}$
and $\hat{V}={\cal H}-{\cal H}_0$. The action of the spin-flip part ${\cal
  H}_{ex;x,y}^{(I)}$ of $\hat{V}$  links to the generalized Green's functions:
$$ G(n,m,n',k, \omega)= \langle k, n' | \hat{G}(\omega) | k, n,
m\rangle
$$ where
%%%%%%%%%%%%%%%%%%%%%%%%%%%%%% EQUATION %%%%%%%%%%%%%%%%%%%%%%%%%%%%%%
\begin{equation}
\label{eq:ElBetwSpinRealSpaceState2} \nonumber
| k, n, m\rangle = \sum_{i} \frac{e^{i k
	(R_{i+\frac{1}{2}}+\frac{n}{2})} }{\sqrt{N}}\crdu{i+\frac{1}{2}}
\crdu{i+\frac{1}{2}+n} \frac{S_{i+m}^{-}}{\sqrt{2 S}} | {\rm FM} \rangle.
\end{equation}
%%%%%%%%%%%%%%%%%%%%%%%%%%%%%%%%%%%%%%%%%%%%%%%%%%%%%%%%%%%%%%%%%%%%%%
These states describe configurations of total momentum $k$ which have
both carriers with spin up and located $n\ge 1$ sites apart, plus a
magnon located anywhere on its sublattice, $ -\infty < m< \infty
$. The conservation of $S_{\text{tot}}^z$ guarantees that states with
two or more magnons cannot appear, so the EOM involve only these two
types of Green's functions.  These equations are listed in Appendix A.

To solve this infinite system of coupled equations, we use two
facts.\cite{FewP} The first is that we can group these states and
their corresponding Green's functions in
terms of an integer index $M$, defined as 
the distance between the two outermost particles. In other words $M=n$
for states $| k, n \rangle$,  while for states $| k, n,
m\rangle$:
\begin{align}
 M= \left \{
\begin{array}{lr}
 n & , 0\leq m \leq n \\ m & , n<m<\infty \\ n-m &, -\infty <m<0
\end{array} \right. .
\end{align}

The importance of this index lies in the fact that $\hat{V}$ only connects
states for which $\Delta M= 0,\pm1$.\cite{note}  We can therefore
group the $3M+1$ Green's functions with the same value of
$M$ into a vector $\mathbf{V}_M$, and rewrite the equations of motions
in matrix form:\cite{FewP}
\begin{align}
\gamma_M \mathbf{V}_M=\alpha_M
\mathbf{V}_{M-1}+\beta_M\mathbf{V}_{M+1}, \label{eq:motionGmatrix}
\end{align}
where the sparse 
matrices $\gamma_M$,  $\alpha_M$ and $\beta_M$ can be read off
directly from the EOM,
Eqs. (\ref{eq:motionGn>0})-(\ref{eq:motionGm}).

To solve this matrix recurrence equation, we use the second useful
observation, namely that since we are working with a finite lifetime
$1/\eta$, the Green's functions must vanish as $M\rightarrow
\infty$.\cite{FewP} As a result, the solution of
Eq. (\ref{eq:motionGmatrix}) must 
have the form:
\begin{align}
 \mathbf{V}_M= \mathbf{A}_M \mathbf{V}_{M-1}, \label{eq:A}
\end{align}
where the continued fraction matrices
\begin{align}
 \mathbf{A}_M= \left [\gamma_M-\beta_M \mathbf{A}_{M+1}\right]^{-1}
 \alpha_M \label{eq:recursive}
\end{align}
are calculated starting from a cutoff $M_c$ for which one sets
\mbox{$\mathbf{A}_{M_c+1} = 0$}. In practice $M_c$ is increased until
convergence within the desired accuracy is reached. While this method
is used here in 1D, it is important to mention that it works for
higher dimensions, as well.\cite{FewP} As an example, in Appendix B we
illustrate how to use this formalism to calculate
$G(0,0,k,\omega)$. Other Green's functions are calculated similarly.

Of course, we could have used basis states with singlet or
triplet-like symmetry, such as $| k,n\rangle_{\mp} \sim | k, n\rangle
\mp| k,-n\rangle$, instead of $|k,n\rangle$ of
Eq. (\ref{eq:ElBetwSpinRealSpaceState1}), etc. The usefulness of these
symmetric states lies in the fact that at $k=0$ the EOM do not mix the
two different symmetries. This reduces the number of equations and
therefore computation times, and furthermore allows one to identify
whether a $k=0$ eigenstate is singlet or triplet-like. 
However, for any $k\ne 0$, the eigenstates do not have a well-defined
symmetry, {\em i.e.} singlet and triplet-like configurations are mixed
together by the EOM. As a result, if interested in the general
solution, there is no advantage in working with symmetric
basis states. 

For completeness, we also mention that even at $k=0$, it is incorrect
to think of these eigenstates as describing singlets or triplets of the charge
carriers. This is because their spins are coupled to the FM background
and cannot be disentangled from it. Even if the charge-carrier part of
these states has singlet or triplet-like symmetry, the full
wavefunction belongs to the $S^z_{tot}=NS$ Hilbert subspace and is,
therefore, not 
a proper singlet or triplet. We continue to use the terms singlet-like or
triplet-like in the following purely for convenience.

Lastly, another advantage of this real-space formulation is that
it allows us to find easily variational solutions for the bound states
(if any exist), and to understand their nature. This is
discussed next. 

\subsubsection{Variational solutions}

The basic idea for the variational solutions (VS) is that if bound
states exist, their wavefunctions decay exponentially with the
distance $n$ between carriers. We should therefore obtain a reasonably
accurate approximation for them if we neglect all terms in the EOM
where the carriers are farther apart than a certain cutoff; this
simplifies the calculation tremendously.  The VS, however, do not
predict correctly the continuum states where the carriers are not
bound to each other. This is not a serious problem, because the
location of such continua can be inferred from the one-carrier
spectra.

We considered two examples of VS, with the carriers
allowed  up to 1 and up to 2 sites away from each other, respectively.
In the following we refer to these as VS1 and VS2.
As the simplest illustration, we discuss here VS1 for model I, in the
triplet-like sector, and then briefly comment on the other cases. 

A triplet-like state allowed in VS1 for model I is:
%%%%%%%%%%%%%%%%%%%%%%%%%%%%%% EQUATION %%%%%%%%%%%%%%%%%%%%%%%%%%%%%%
\begin{equation}
\EqLabel{vs1}
| k,1 \rangle_+ = \sum_{i}\frac{e^{ik R_{{i}}}}{\sqrt{N}}
\frac{\crdu{{i}-{1\over 2}}\crdd{{i}+{1\over 2}}+\crdd{{i}-{1\over
	  2}}\crdu{{i}+{1\over 2}}}{\sqrt{2}}
| {\rm {\rm FM} }
\rangle.
\end{equation}
%%%%%%%%%%%%%%%%%%%%%%%%%%%%%%%%%%%%%%%%%%%%%%%%%%%%%%%%%%%%%%%%%%%%%%
Exchange couples it  to the one-magnon state:
%%%%%%%%%%%%%%%%%%%%%%%%%%%%%% EQUATION %%%%%%%%%%%%%%%%%%%%%%%%%%%%%%
\begin{equation}
\EqLabel{vs2}
| k,1,1 \rangle_+ = \sum_{i}\frac{e^{ik R_{{i}}}}{\sqrt{N}}
\crdu{{i}-{1\over 2}}\crdu{{i}+{1\over 2}}
 \frac{S_{i}^-}{\sqrt{2S}} | {\rm FM} \rangle
\end{equation}
%%%%%%%%%%%%%%%%%%%%%%%%%%%%%%%%%%%%%%%%%%%%%%%%%%%%%%%%%%%%%%%%%%%%%%
where the magnon is located directly between the two carriers. Of course, the
magnon is free to move so this will further link to the states  $|
k,1,m\rangle_+ = \sum_{i}\frac{e^{ik R_{{i}}}}{\sqrt{N}} 
\crdu{{i}-{1\over 2}}\crdu{{i}+{1\over 2}}
\frac{S_{i+1-m}^-+ S_{i-1+m}^-}{2\sqrt{S}} | {\rm FM} \rangle$  for any
$m\le 0$. Note that the basis state of Eq. (\ref{vs1}) is also linked
by spin-flip processes directly to $|k,1,m=0\rangle_+$.

The EOM for the  Green's functions associated with these (very few)
basis states are listed and solved in Appendix C. They have 3
poles, {\em i.e.} 3 triplet-like bound states are predicted by this
approximation. Their energies (for $J\rightarrow 0$, to simplify their
expressions) are:
\begin{align}
 &\omega_1^+ = -\frac{1}{4} J_0 \left(3-4 S+\sqrt{1+16 S^2}\right)
 \label{eq:ElBetwSpinVariationalGS}, \\ &\omega_2^+ = \frac{1}{4} J_0
 \left(-3+4 S+\sqrt{1+16 S^2}\right), \\ &\omega_3^+ =
 2J_0S. \label{eq:ElBetwSpinVariationalCont}
\end{align} 
As shown in the next section, the third pole falls inside a continuum,
in other words it does not survive as a bound state if we relax the
variational approximation. The other two, however, are outside the
expected continua and indeed will be shown to give a reasonable
approximation for the energies of the two triplet-like bound states
predicted by the exact solutions for this model.

While this is a nice example of the usefulness of the VS,
some care is needed. For example, VS1 predicts states with
triplet-like symmetry for all $k$.  This contradicts our previous statement
that the exact solution only finds states with well defined symmetry
at $k=0$. The reason for this is trivial: no carrier hopping is
possible within the triplet-like sector of VS1, since there is no ``on-site
triplet'', and states with $n\ge 2$ are not allowed. (This also
explains why the energies of Eqs.
(\ref{eq:ElBetwSpinVariationalGS})-(\ref{eq:ElBetwSpinVariationalCont})
do not depend on $t$ or $U$). This discrepancy is solved by VS2
where the EOM couple the singlet and triplet-like sectors for all $k\ne 0$,
and indeed we find that the bound states have well-defined
singlet/triplet-like symmetry only at $k=0$. This is true for both models I
and II.

VS1 in the singlet sector works similarly, and also predicts 3 bound
states for model I. Model II is predicted to have only one
singlet-like bound state and no triplet-like bound states. This
already suggests that their physics may be rather different. Finally,
we note that for VS2 we also restricted the distance between the
magnon and its closest carrier to be up to 2 sites. This is because
the magnon propagation is controlled by $J$, which is the smallest
energy scale. Further details and more VS results are available in
Ref. \onlinecite{Mirko}.

\section{Results} \label{sec:Results}

\subsection{Single spin-down charge carrier}

Since the case of a single spin-up carrier is trivial, with a single
eigenstate of energy $E_{k\uparrow}$, here we discuss only the single
spin-down carrier case. If the spin-flip processes were turned off,
$\mathcal{H}_{ex;x,y}^{(I,II)}\rightarrow 0$, eigenstates with a total
momentum $k$ and $S_{\text{tot}}^z=NS-{1\over 2} $ would either have
the carrier with spin down and energy $E_{k\downarrow}$, or the
carrier with spin up and energy $E_{k-q,\uparrow}$ while the remaining
momentum and spin is carried by a magnon (spin wave) of energy
$\Omega_q$. These latter states give rise to a continuum spanning the
interval $\{ E_{k-q,\uparrow} + \Omega_q\}_q$. Spin-flip processes
hybridize the states in the continuum with the discrete spin-down
carrier state. Of course, the continuum keeps its location, but the
discrete state is pushed even lower (for $J_0>0$) and gives rise to a
low-energy, infinitely lived quasiparticle: the spin polaron.

\begin{figure}[t]
\includegraphics[angle=-90,width=.95\columnwidth]{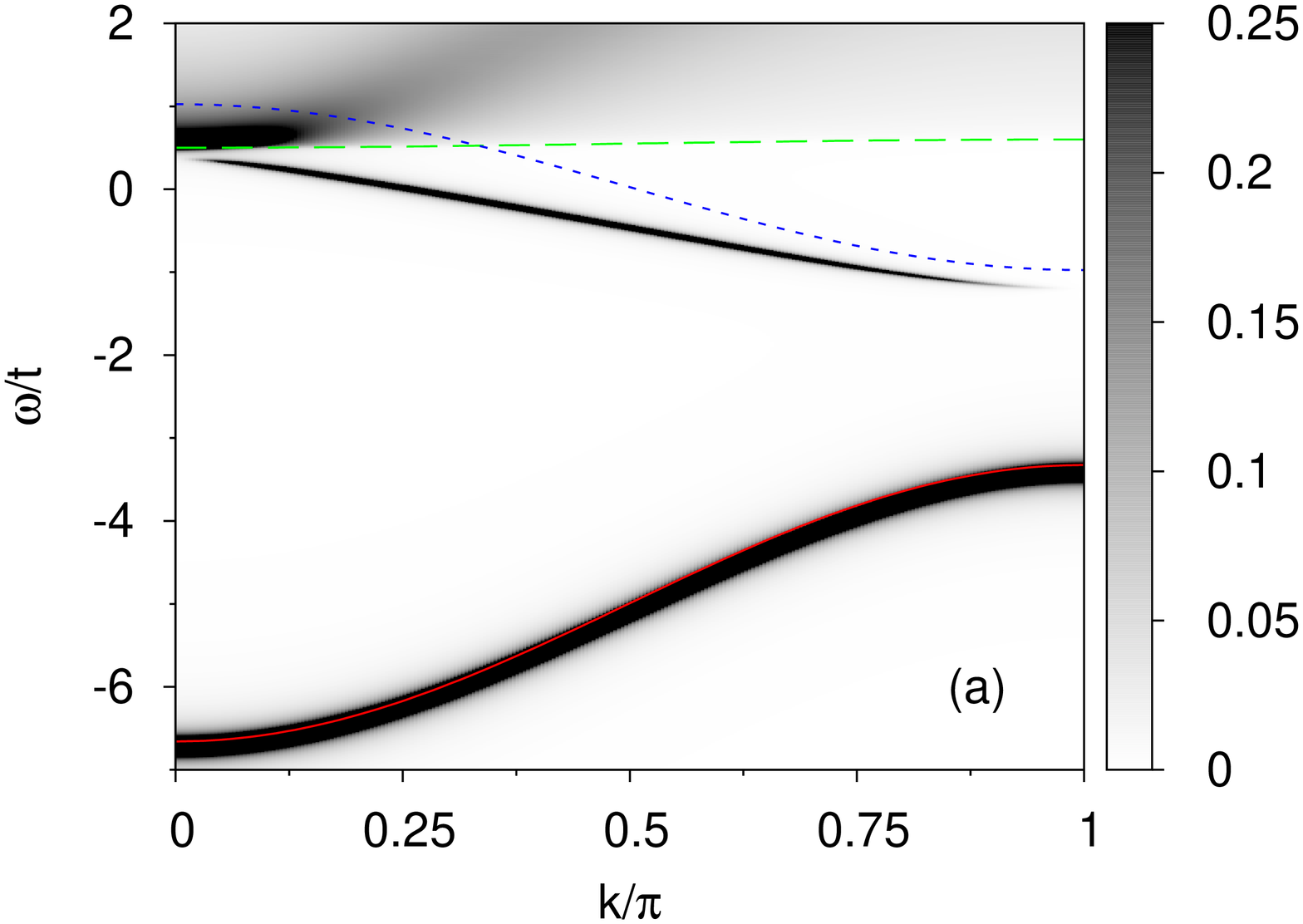}
\includegraphics[angle=-90,width=.95\columnwidth]{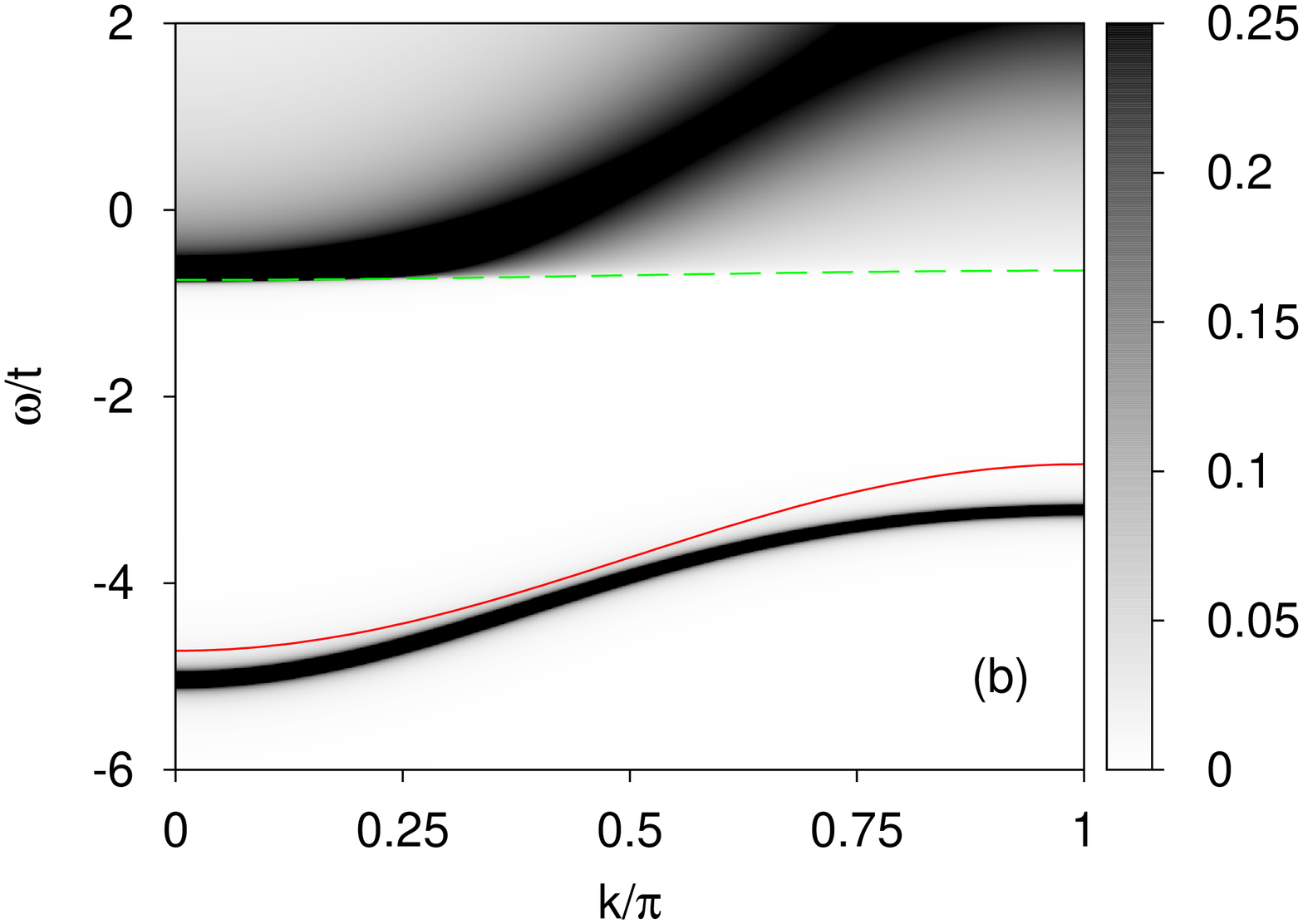}
\caption{(Color online) Contour plot of the DOS
  $\rho_{\downarrow}(k,\omega)$ for (a) 
model I and (b) model II. The full red lines are the approximations
given by Eq. (\ref{eq:spModelII}) and (\ref{eq:spModelI}) and the
dashed green lines show the predicted onset of the continuum at
$\text{min}_q(E_{k-q,\uparrow}+\Omega_q)$. The dotted blue line in
panel (a) shows
the approximation of Eq. (\ref{eq:sp*ModelI}). Parameters are
$J_0=5t$, $J=0.05t$, $S=\frac{1}{2}$ and $\eta=0.01t$ in (a),
$\eta=0.02t$ in (b).}
\label{fig3}
\end{figure}

These general expectations are confirmed in Fig. \ref{fig3}, which
shows contour plots of the density of states (DOS)
$\rho_{\downarrow}(k,\omega)=-\frac{1}{\pi} \text{Im}
G_{\downarrow}(k,\omega)$ for models I and II, for $J_0=5t$,
$J=0.05t$, $S=\frac{1}{2}$. We only show the lower part of the
continuum, which indeed begins at
$\text{min}_q(E_{k-q,\uparrow}+\Omega_q)$ (dashed green line), as
expected. The low-energy states of both spectra are the discrete states
associated with the spin-polarons. For $J_0 < 0$ these discrete states
appear above the continuum, so in that situation the low-energy
physics is described by incoherent continuum states.

The nature of the spin-polaron can be understood by considering the
limit $J_0 \gg t,J$. We start with model II. The interaction
$\mathcal{H}_{ex}^{(II)}$ is minimized by an on-site singlet between
the carrier and its spin:
\begin{align}
  | s \rangle_i= \frac{1}{\sqrt{2 S+1}} \left \{ \sqrt{2S}
  \crdd{i}-\frac{1}{\sqrt{2S}} \crdu{i} S_i^- \right \} | {\rm FM}
  \rangle,
  \label{eq:SameSitespState}
\end{align}
with eigenenergy $-\frac{J_0}{2}(S+1)$. Strictly speaking, this is a
singlet only for $S={1\over 2}$, but we will use this term in the
following.  Hopping $\hat{T}$ lifts the degeneracy, and to first
order in $t$ and $J$ the spin-polaron energy is given by:
\begin{align}
 E_P^{(II)}(k)=-\frac{J_0}{2}(S+1)+\frac{2S}{2S+1}\epsilon(k)+\frac{2JS}{2S+1}.
 \label{eq:spModelII} 
\end{align} 
The spin-polaron state $| P_{II},k\rangle =\frac{1}{\sqrt{N}}\sum_i
e^{ikR_i}$ \mbox{$| s \rangle_i$} thus describes a  singlet, or
bound state between the carrier and a magnon at the same site, which
propagates with an effective hopping amplitude $2St/(2S+1)<t$,
suppressed because of magnon cloud overlap. The last term is the FM
exchange energy lost in the magnon's presence. The full (red) line in
Fig. \ref{fig3}(b) shows this variational approximation, which is
already quite accurate even for $J_0=5t$. In fact,
Eq. (\ref{eq:spModelII}) gives a reasonable approximation for the
$k\approx 0$ part of the polaron dispersion down to $J_0\sim t$,
showing that this picture of the polaron as a local singlet between the
carrier and its spin is quite robust. What happens as $J_0$
decreases is that the polaron band moves closer to the continuum, and
the $k\rightarrow \pi$ part of its dispersion ``flattens out'' below
the continuum, in typical polaronic fashion (for more details, see
Ref. \onlinecite{FM}).

Similar considerations apply to model I. Here,
$\mathcal{H}_{ex}^{(I)}$ is minimized by the three-spin polaron
(3SP):\cite{3SP}
\begin{align}
| 3SP \rangle_{i+\frac{1}{2}} = \sqrt{4S \over 4S+1}\left \{
\crdd{i+\frac{1}{2}}- \crdu{i+\frac{1}{2}}
\frac{S_i^-+S_{i+1}^{-}}{4S} \right \}| {\rm FM} \rangle \nonumber
\end{align}
 with eigenenergy $-J_0(S+\frac{1}{2})$. It describes a bound state
 between the carrier and a magnon on either of the neighbor spin sites. Hopping
 lifts the degeneracy and, to first order in $t$ and $J$, the energy
 of the polaron eigenstate $| P_I, k\rangle=\frac{1}{\sqrt{N}} \sum_i e^{ik
 R_{i+\frac{1}{2}}} | 3SP \rangle_{i+\frac{1}{2}}$ is:
\begin{align}
 & E_{P}^{(I)}(k)=
 -J_0\left(S+\frac{1}{2}\right)+\epsilon(k)\frac{4S+{1\over2}}
 {4S+1}+\frac{JS}{4S+1}.  
 \label{eq:spModelI} 
\end{align}
This expression is shown as a full (red) line in Fig. \ref{fig3}(a),
and is in very good agreement with the exact dispersion already for
$J_0=5t$. The agreement remains reasonable down to $J_0=t$ at
$k\approx 0$, indicating that the description of this spin-polaron as
a propagating 3SP is also robust.

For sufficiently large $J_0$ values, model I has a second discrete
state below the continuum. This is based on an excited state of
$\mathcal{H}_{ex}^{(I)}$. In the $J_0 \gg t,J$ limit,  it leads to $|
P_{I}^*,k \rangle=\frac{1}{\sqrt{4SN}}\sum_i
e^{ikR_{i+\frac{1}{2}}}\crdu{i+\frac{1}{2}} (S_i^--S_{i+1}^{-}) | {\rm
FM} \rangle$, with energy:
\begin{align}
 E_{P^*}^{(I)}(k)=\frac{J_0}{2}(2S-1)-\frac{1}{2}\epsilon(k)+JS.
 \label{eq:sp*ModelI} 
\end{align}
The agreement between this approximation (thin blue line) and the
exact result shown in Fig. \ref{fig3}(a) is rather poor, especially
close to the continuum, but it improves with increasing $J_0$. More
details on these additional states are discussed in
Ref. \onlinecite{FM}, in a 2D context. 

To summarize the results so far, in the $S_{\text{tot}}^z=NS-{1\over
2} $, one-carrier sector, all three models have a low-energy,
infinitely lived quasiparticle. The mapping from model II to model
III is straightforward, since the polaron of model II has a
singlet-like core, precisely like the ``hole'' of model III. Of
course, a proper mapping requires some rescaling of the parameters:
the ``hole'' hopping integral is different from that of the charge
carrier itself. The discussion above identified the origin of
this rescaling (polaron cloud overlap) and the expected renormalization.

At first sight, the polarons of model I and II seem to be quite
different. However, we can rewrite $$|
P_{I},k\rangle=\sum_i \frac{e^{ikR_i}}{\sqrt{2N}} \left (\sqrt{2S}
d_{k,i,\downarrow}^{\dagger}-\frac{1}{\sqrt{2S}}
d_{k,i,\uparrow}^{\dagger}S_i^- \right)| {\rm FM} \rangle,$$
{\em i.e.} as a  singlet between the carrier in an ``on-site''
orbital $d_{k,i,\sigma}^\dagger= \frac{1}{\sqrt{4S+1}} \left
(e^{i\frac{k}{2}} c_{i+\frac{1}{2},\sigma}^{\dagger}+e^{-i\frac{k}{2}}
c_{i-\frac{1}{2},\sigma}^{\dagger} \right )$, and its local spin. This
is exactly like the polaron of model II if we replace $d_{k,i,\sigma}
\rightarrow c_{i,\sigma}$. 

These $d_{k,i,\sigma}$ operators are similar to the $P_{i,\sigma}^{(S)}$
operators in the Zhang-Rice approach: they are linear combinations of
carrier orbitals at sites neighboring a spin-site, and they are not
orthogonal to one another. The main difference is that our linear
combinations have $k$-dependent phases. This allows us to avoid the
need for a diverging normalization factor $\beta_{\bf k}$ in $|
P_{I},k\rangle$, which then lead to the introduction of the much
more complicated orbitals $\phi_{i\sigma}$ in the ZRS.\cite{ZR} While the
discussion here is for 1D, similar considerations hold in 2D.\cite{FM}

What we found so far, then, is that all three models have the same
low-energy physics in the non-trivial single-carrier sector: the
low-energy state is an infinitely lived quasiparticle in all cases,
with a polaron core similar to a ZRS.
This suggests that we can indeed model the low-energy physics of the
two-sublattice, two-band Hamiltonian with a much simpler $t$-$J$ model
for ``holes'', similar to Zhang and Rice's mapping of the
two-sublattice, three-band Emery model to a $t$-$J$ model for ZRS.
The next question is whether this mapping also describes well the
interactions between quasiparticles.

\subsection{Two charge carriers in the $S^z_{\rm tot}=NS$ sector}

As already mentioned, if both carriers have spin up ($S^z_{\rm
tot}=NS+1$), there are no interactions between them. If both are
injected with spin down ($S^z_{\rm tot}=NS-1$), the magnon-mediated
interactions turn out to be weak or repulsive and there is no qualitative
difference between the  low-energy spectra of these models, as we
discuss at the end of this section.

The interesting case is the $S^z_{\rm tot}=NS$ sector, {\em i.e.} when
one carrier is injected with spin up and the other one with spin down. As we
show now, here we can identify qualitative differences between models
I and II (model III cannot be used here, since it does not
distinguish between a spin-up carrier and a lattice spin).

\begin{figure}[t]
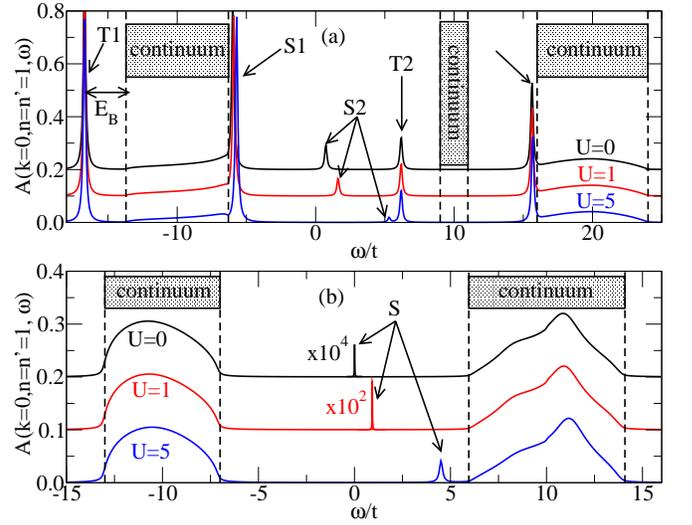

 \includegraphics[width=\columnwidth]{fig4a.eps}
 \includegraphics[width=\columnwidth]{fig4b.eps}
 \caption{(Color online) Spectral function $A(k=0,n=n'=1,\omega)$ for
 model I (a) 
 and model II (b). Dashed lines mark the expected continuum
 boundaries. Discrete states, {\em i.e.} bipolarons, are labeled
 according to their symmetry, S for singlet-like and T for
 triplet-like. Parameters are $J_0=20t$, $J=0.05t$, $S=1/2$,
 $\eta=0.1t$ and $U/t= 0, 1, 5$.}
  \label{fig4}
\end{figure}

\begin{figure}[b]
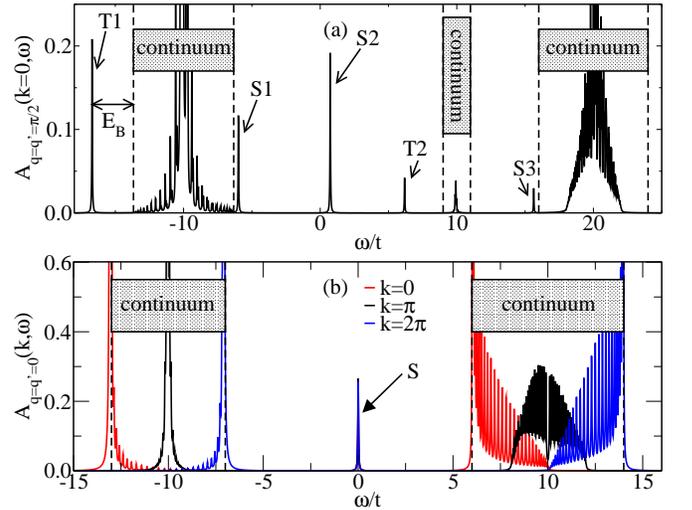

 \includegraphics[width=\columnwidth]{fig5a.eps}
 \includegraphics[width=\columnwidth]{fig5b.eps}
\caption{ (Color online) (a)
  $A_{q=q'=\frac{\pi}{2}}(k=0,\omega)$ 
for model I.  Using $q=q'=\frac{\pi}{2}$ ensures that states
with both symmetries have non-vanishing weight. (b) $A_{q=q'=0}(k,
\omega)$ for model II and $k=0,\pi,2 \pi$. Dashed lines mark the
  expected continuum 
boundaries. Discrete states, {\em i.e.} bipolarons, are labeled
according to their symmetry, S for singlet-like and T for
triplet-like. Parameters are $J_0/t=20$,
$J/t=0.05$, $S=1/2$, $U/t= 0$, $\eta/t=0.02$ and $N=60$ for (a), $N=50$ for (b).}
  \label{fig5}
\end{figure}

$A(k=0,n=n'=1,\omega)=-\frac{1}{\pi} \text{Im} G(n=n'=1,k=0,\omega)$
obtained with the real-space solution for both models are shown in
Fig. \ref{fig4}. Their counterparts
$A_{q,q'}(k=0,\omega)=-\frac{1}{\pi} \text{Im}G(k=0,q,q',\omega)$, are
shown in Fig. \ref{fig5}. Since both methods are exact, these spectral
weights must have poles at exactly the same eigenenergies, however the
weight at these poles differs since we are projecting on different
types of states. In both cases spectra are displayed for $J_0/t=20$,
$J/t=0.05$ and $S=\frac{1}{2}$. The value of $J_0$ was chosen so large
intentionally in order to spread out the different features of the
spectrum.

In the absence of interactions, the spectrum for two carriers equals
the convolution of the corresponding one-particle spectra, which are
known. This is why we expect to see a high energy continuum describing
scattering states consisting of two spin-up carriers and a magnon,
spanning energies
$\{E_{q,\uparrow}+E_{q',\uparrow}+\Omega_{k-q-q'}\}_{q,q'}$, arising
from the convolution between the high-energy continuum in the
spin-down carrier spectrum and the spin-up carrier eigenstate.
Neglecting a small $J$ term, the edges of this continuum are at
$\pm4t+ 2 \gamma J_0S$, {\em i.e.} $[16t,24t]$ for model I and $[6t,
14t]$ for model II, for these parameters. These values are marked by
the dashed vertical lines.

The real-space solution indeed shows a high-energy continuum for both
models at precisely the expected locations. However, the k-space
solution (Fig. \ref{fig5}(a)) has finite weight only in part of the
continuum. The reason for this is that here we project on delocalized
states where each carrier has a well defined momentum $\kh \pm
q$. Depending on $k$ and $q$, sections of the continuum have vanishing
overlap with these states. This is confirmed in Fig. \ref{fig5}(b),
where $A_{q=q'=0}(k,\omega)$ is shown for $k=0, \pi, 2\pi$. Taken
together, the three curves cover the full interval where the continuum
is expected. Note that the spectrum for $k=2 \pi$ is different from
that for $k=0$, since for the former we project on states $\ckdu{\pi}
\ckdd{\pi} | {\rm FM} \rangle$, whereas for the latter we project on
$\ckdu{0} \ckdd{0} | {\rm FM} \rangle$.

The k-space solution also has oscillating weight inside the continua. This
is a finite-size effect due to the relatively small value of $N=60$
unit cells, and can be fixed using a larger $N$. However, since here we
project on delocalized states, an increase in $N$ results in a
smaller weight for all bound states, as already discussed. Therefore,
for our purposes  it is not beneficial to use a larger $N$.

Next, the convolution of the polaron state and the spin-up state must
result in a low-energy continuum
$\{E_{P}^{(I,II)}(q)+E_{k-q,\uparrow}\}_q$. Using the asymptotic
expressions from Eqs. (\ref{eq:spModelII}) and (\ref{eq:spModelI}) for
the polaron energies and ignoring the small $J$ term, for $k=0$ the
edges of this continuum are at $\pm 2t \left(1
+\frac{4S+{1\over2}}{4S+1}\right) - \frac{J_0}{2}$ for model I and at
$\pm 2t \left(1 +\frac{2S}{2S+1}\right) - \frac{J_0}{2}$ for model II,
respectively.  For the parameters of Figs. \ref{fig4} and \ref{fig5}
this corresponds to $[-13.7t, -6.3t]$ and $[-13t,-7t]$,
respectively. These values are marked by vertical dashed lines and are
also in excellent agreement with the low-energy continua found by the
exact solutions (similar considerations to those discussed above apply
to the k-space solution).

For large values of $J_0/t$ the spin-down carrier spectrum of model I
has the additional discrete state between the polaron band and the
continuum ({\em cf.}  Fig. \ref{fig3}). This should also give rise to
a continuum $\{E_{P^*}^{(I)}(q)+ E_{k-q,\uparrow}\}_q$, which
corresponds to $[9t,11t]$ for these parameters. This continuum is
absent in the real-space solution. This is not surprising because the
asymptotic expression for $|P_I^*,k\rangle$ has zero overlap with any
state with a spin-up and a spin-down carrier like those used in the
projection. In reality, there are small corrections to this asymptotic
expression and indeed, the k-space solution shows a (restricted)
continuum here, with a weight that is indeed orders of magnitude
smaller than for the other two continua.

We conclude that there is excellent agreement between the continua
displayed by the exact solutions, and those expected based on the
convolutions of the corresponding one-carrier spectra; this is a good
consistency check.

If there are sufficiently strong interactions between carriers so as
to bind them in pairs, this should result in discrete states appearing
in the spectrum outside these continua.  The exact
solutions show five such states for model I, and one for model II.

\begin{figure}[t]
\includegraphics[width=\columnwidth]{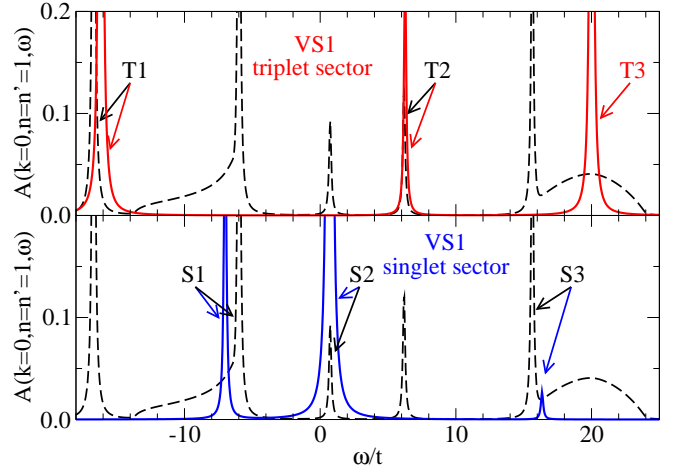}
\caption{(Color online) Exact spectral weight $A(k=0,n=n'=1,\omega)$
  (dashed black line) and the corresponding VS1 predictions in the
  triplet-like (top panel, full red line) and singlet-like (bottom panel, full
  blue line), for model I. Parameters are $J_0/t=20, J/t=0.05,
  U/t=0, S=1/2, \eta/t=0.1$.}
\label{fig6}
\end{figure}

We have checked that these are indeed discrete states (as opposed to
very narrow continua) by verifying that they are Lorentzians of width
$\eta$.\cite{Mirko} This finite width is responsible for the apparent
overlaps of S1 and S3 with their nearby continua, but they become
distinct features as $\eta \rightarrow 0$. The nature of these states
is further confirmed in Fig. \ref{fig6}, where we compare the exact
spectral weights to the predictions of  VS1
for the triplet-like (top panel) and singlet-like (bottom panel) sectors.  Eqs.
(\ref{eq:ElBetwSpinVariationalGS})-(\ref{eq:ElBetwSpinVariationalCont})
predict three peaks in the triplet-like sector. The lower two are in good
agreement with two of the exact peaks; the third falls in the center
of a continuum, and is therefore not indicative of a true
bound-state. In the singlet-like sector, VS1 predicts three peaks as
well. The lowest and highest of these happen to fall at the edge of
continua. A more accurate approximation like VS2 shows that
they are pushed just outside those continua, where indeed there are
discrete states in the exact solution. The central peak also agrees
well with a discrete peak in the exact solution. For model II, VS1
finds one singlet peak, in agreement with the peak appearing near
$\omega\sim U$ in the exact solution.\cite{Mirko}

Since the variational solutions include only states where the carriers
are confined to be close to one another, the good agreement with the
discrete states of the exact solutions validates the latters'
identification as bound states, {\em i.e.} bipolarons.  Their 
triplet-like or singlet-like symmetry at $k=0$ is also in perfect
agreement with what 
is inferred from the exact solution by looking at the relative phases
of the $n=1$ and $n=-1$ Green's functions, for example. As 
mentioned, at $k\ne 0$ these states no longer have a definite
symmetry.

From the VS or from the dependence on $n$ of $G(n, n', k,\omega)$ at
energies $\omega$ close to one of these peaks, we can infer the
underlying nature of these polarons. For example, the bound state of
model II has most of its weight on the configuration where both
carriers are on the same site, {\em i.e.} $| k=0, n=0\rangle$.  This
explains the small spectral weight for this state in
Fig. \ref{fig4}(b), where we project on $| k=0, n=1 \rangle$ states
whose contribution is exponentially smaller, especially for smaller
$U$. It also explains the strong dependence of its energy on $U$:
having both carriers on the same site costs $U$. Finally, it also
explains why the energy of this state hardly depends on $J_0$: the
carriers form a singlet with each other when on the same site, and
there is no exchange between this singlet and the lattice spins. As a
result, this state of energy $\omega \sim U$ is always located above
the low-energy spin-up carrier+polaron continuum, whose energy is
$\sim -J_0/2$ lower because of exchange between carriers and spins.
This state (which is very similar to S2 of model I) has a
non-monotonic dispersion with $k$ in certain regions of the parameter
space.\cite{Mirko} However, it is rather uninteresting overall, since
it is not bound by magnon-mediated interactions and,  moreover, it is a
high-energy state.

In contrast, model I has a triplet-like bipolaron (T1) as its low-energy
eigenstate, in addition to four other higher energy bipolaron
states. While we can analyze and understand the nature of all these
higher-energy bipolarons, in the following we focus on the low-energy
bound state  which is of primary interest to us.

\begin{figure}[t]
 \includegraphics[angle=0,width=\columnwidth]{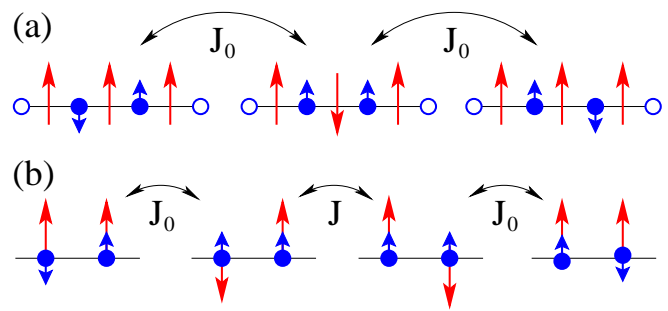}
  \caption{(Color online) (a) The three highest-weight configurations
	of the low-energy T1 bipolaron of model I, illustrating the
	magnon-mediated interaction that binds the carriers.  There are
	matrix elements of order $J_0$ between these states, due to
	carrier-spin spin-flip processes. (b) A possible analogue in model
	II for the magnon-exchange process shown in (a). The central
	matrix element is now of order $J$.  Symbols have the same meaning
	as in Fig. \ref{fig2}.}
  \label{fig7}
\end{figure}

\subsection{The low-energy bipolaron of model I}

The three configurations with the highest-weight contributions to this
bipolaron's wavefunction are depicted in Fig. \ref{fig7}(a). They
clearly illustrate the underlying mechanism responsible for binding
the two carriers as being due to the back-and-forth exchange of a
magnon, which is emitted by the spin-down carrier and then absorbed by
the other, initially spin-up carrier. This process is facilitated by
the fact that the carriers sit on adjacent sites and {\em
simultaneously} interact with the lattice spin located between
them. The triplet-like symmetry at $k=0$ is a consequence of the fact that
the symmetric combination (bonding state) has the lowest energy, all
the more so since it also avoids the cost of the on-site repulsion
$U$. Indeed, Fig. \ref{fig4}(a) confirms that the energy of this state
is independent of $U$.

In Fig. \ref{fig7}(b) we sketch what would be the analog in model II
for this magnon-mediated interaction. Because here each carrier
interacts only with its own spin, the magnon has to hop from one site
to the other. This is a process controlled by $J$, which is the
smallest energy scale. As such, it is not sufficient to bind the two
carriers together, and indeed we find no such bound states in the
low-energy spectrum of model II. Clearly, having
carriers and spins on different sublattices greatly facilitates the
magnon-mediated interactions in model I.

\begin{figure}[b]
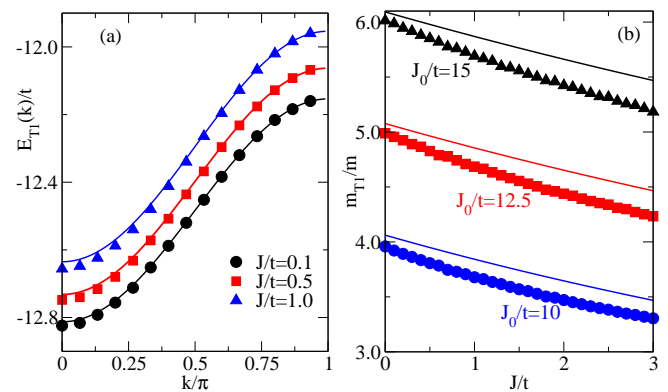

 \includegraphics[width=.49\columnwidth]{fig8a.eps}
 \includegraphics[width=.49\columnwidth]{fig8b.eps}
  \caption{(Color online) (a) The dispersion of the low-energy T1
  bipolaron of model 
  I, for $J_0/t=15$ and $J/t=0.1,0.5,1$.  (b) Bipolaron
  mass in units of the free carrier mass, $m_{T1}/m$, as a function of $J/t$
  for $J_0/t=10, 12.5, 15$. Other parameters are $S=\frac{1}{2}$,
  $\eta /t=0.02$. Symbols are the exact results obtained with the
  real-space solution, while solid lines give a perturbative
  approximation discussed in the text.}
  \label{fig8}
\end{figure}

The dispersion $E_{\text{T1}}(k)$ of this low-energy bipolaron is
plotted in Fig. \ref{fig8}(a) for three values of $J$. Changing $J$
shifts the curves vertically, as expected since the FM exchange energy
lost in the presence of the magnon is increased. However, the shape of
these curves does not change much, confirming the fact that the
binding and dynamics of the bipolaron are primarily controlled by
$J_0$ and $t$. The dispersion is consistent with nearest-neighbor
hopping, but with a renormalized effective mass.

The bipolaron mass expressed in terms of the free carrier mass is
shown in Fig. \ref{fig8}(b) vs. $J/t$, for several values of
$J_0/t$. These bipolarons are remarkably light, with $m_{T1}/m\approx
4$ for $J_0/t=10$ and further decreasing with decreasing $J_0$ and/or
increasing $J$.

\begin{figure}[t]
\includegraphics[width=0.8\columnwidth]{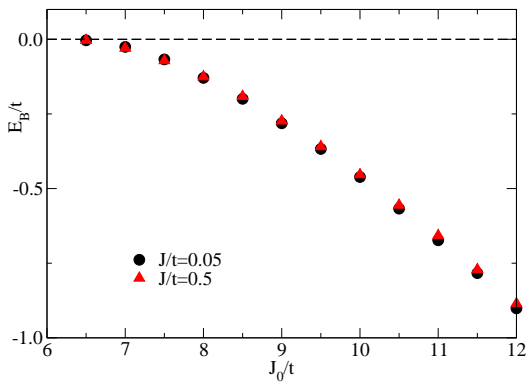}
\caption{Bipolaron binding energy $E_B$ as a function of $J_0$ for
two values of $J$. These  exact values are for $S={1\over2}$, and are
independent of $U$. The bipolaron unbinds if $J_0/t < 6.5t$. }
\label{fig9}
\end{figure}

To understand these trends, we used  Rayleigh-Schr\"odinger
perturbation theory to third order in $t$ and $J$.\cite{Sakurai}
Details are summarized in Appendix D. The results are shown by the
full lines in Fig. \ref{fig8}, and are in reasonable agreement with
the exact results. In particular, for $S={1\over 2}$ we find the
bipolaron effective mass to be:
\begin{align}
\frac{m_{\text{T1}}}{m} =& \frac{0.406 J_0^2/t}{J_0 + 0.570 J}\approx
 0.406 \frac{J_0}{t}-0.231 \frac{J}{t}. \label{eq:T1dispersion}
\end{align}
The dependence on $J_0$ and $J$, which agrees with that displayed by
the exact results, can be understood as follows. Consider moving the 3
configurations shown in Fig. \ref{fig7}(a) by one site to the
right. For the left and the right configurations, this can be done
simply with two consecutive hops of the two charge carriers. However,
for the central configuration, hopping the first carrier to the right
costs an energy of order $J_0$, since it leaves the vicinity of the
magnon. Hopping the second carrier then leaves the magnon
``behind''. This can be fixed either by absorbing the magnon first and thus
switching to one of the other two configurations, or, for large enough
$J$, the magnon has a high probability to hop on its own. Of course,
in reality things are more complicated, but this analysis gives a
rough idea why an increase in $J$ and/or decrease in $J_0$ makes the
bipolaron more mobile.

So far, results have been shown for values of $J_0/t$ that are rather
unphysically large. The final question to address is what happens for
smaller values of $J_0/t$. The answer is provided in Fig. \ref{fig9},
where we plot the evolution of the bipolaron binding energy $E_B$
(measured with respect to the lower edge of the low-energy continuum,
see Figs. \ref{fig4}(a) and \ref{fig5}(a)) with $J_0/t$. For negative
values of $E_B$ the bipolaron is below the continuum and therefore
stable, while $E_B\rightarrow0$ means that the magnon-mediated
interaction is no longer sufficient to bind the bipolaron and instead
it dissociates into the continuum of states with a polaron plus a
spin-up carrier.

Fig. \ref{fig9} shows that the bipolaron is stable only for $J_0/t
\gtrsim 6.5$, roughly independent of the value of $J$ and certainly
independent of the value of $U$. This is consistent with the fact that
$J_0$ controls the magnon-mediated interaction in model I, and that
this interaction must be sufficiently large to compensate for the loss
of kinetic energy in order to bind the carriers.

An in-depth analysis of the significance of these results is offered in
the last section. Here we conclude  by noting the
extreme importance of the lattice structure to the low-energy
physics in the two-carrier sector. Because in model I the two carriers
can simultaneously interact with the same spin, the magnon-mediated
exchange is very efficient and results in a strong effective
attraction that can stabilize a bipolaron. In model II, a spin can
interact at most with one carrier at a time. As a result, the
magnon-mediated interaction is very weak and low-energy bipolarons do
not form. The two models are no longer qualitatively similar in this
two-carrier sector.

\subsection{Two spin-down carriers}

It is possible to obtain the exact $T=0$ solution in the two-carrier,
$S^z_{\rm tot}=NS-1$ subspace with the real-space formalism. The only
complication is that spin-flip processes link the two spin-down
carrier states  both to states with one magnon and one spin-up carrier,
and to states with two magnons and both carriers with 
spin-up. Thus, we can now have up to four particles (carriers or
magnons) in the system, but this can be handled by the formalism
introduced in Ref. \onlinecite{FewP}.

However, perturbational arguments described next show that it is
always energetically unfavorable for a low-energy bipolaron to form in this
sector. The lowest energy feature is, therefore, the two-polaron
continuum whose location is already known to be $\{E_P^{(I,II)}(k-q) +
E_P^{(I,II)}(q)\}_q$, and the full calculation becomes unnecessary. In
model III, the situation is qualitatively similar since no
interactions appear between two ``holes''.

Consider model I in the limit $t\rightarrow 0, J\rightarrow0$. If the
two carriers are two or more sites apart then each forms a 3SP, and
the total energy is $2 E_{\text{P}}^{(I)}\rightarrow -J_0(2S+1)$. If
the carriers are on neighboring sites, simultaneous interactions with
the central spin change this energy to a value that is calculated easily by
exact diagonalization. We find\cite{Mirko} that this energy is {\em
larger} than $2 E_{\text{P}}^{(I)}$ for any $S$ and $J_0$, in other
words it is energetically favorable to keep the 3SP spatially
apart. Thus, in model I the magnon-mediated interactions are
repulsive in this spin-sector. In model II, these interactions are very
weak because each charge carrier interacts only with its own spin/magnon.
Turning on the hopping will further favor unbound polarons, since
they are lighter and therefore can lower their kinetic energy more
than a bipolaron. 

This is why we do not expect low-energy bound states in any of these models, in
this sector, so we do not consider it any further.

\section{Summary and discussion} \label{sec:Summary}

In this article we used exact, $T=0$ solutions for one and two carriers
doped in a ferromagnetic background to compare the low-energy physics
of three different models. The goal was to understand if the simpler
models describe the same low-energy physics as the more complex one.

In the single-carrier sector this turned out to be the case: all three
models have a low-energy polaron whose core can be thought of as a
singlet between the carrier and a local spin. In the 
two-sublattice model I, a 3SP description is more appropriate but it
can be reduced to a propagating singlet using a construction somewhat
similar to the ZRS. Based on this, one could conclude that a simple
one-band model provides an appropriate description for carriers in a
FM background.

However, results from the two-carrier sector paint a different
picture, especially when the carriers are injected with opposite
spins. Here, for the two-sublattice model we find that the ability
of both carriers to interact simultaneously with a common spin leads to an
attractive, magnon-mediated interaction, of order $J_0$. Such an
interaction is evidently not included in a $t$-$J$ model, where the
$J_0$ energy scale was integrated out. In fact, for the FM background
we cannot even use the $t$-$J$ model to describe this case, because
``holes'' only appear when doping with spin-down carriers, and there
is no counterpart for a spin-up carrier. This is why here we studied
the intermediate model II, where we know that a
spin-down carrier can form its singlet-like ``hole'', but we can also
treat spin-up carriers. The fact that in model II only one carrier can
interact with a given spin, however, leads to a very different mechanism for
the magnon-mediated interaction. In particular, the strong attraction
that appears in model I does not arise in model II.
 
Similar differences appear in the two spin-down sector. We argued that
here each carrier forms its polaron and they move independently, so
the ``big picture'' seems to be the same in all the models: none gives
rise to a strong-enough magnon-mediated attraction to be able to bind
the polarons. However, there are differences in details. As discussed
in the last subsection, for model I here we expect a local repulsion
between polarons, again controlled by $J_0$, due to the additional
exchange cost when the two spin-down carriers are forced on
neighboring sites and frustrate each other in forming their 3SP
clouds. In contrast, in model II each spin-down carrier can form a
singlet with its spin, and there is no difference if they are
neighbors or not; hence, there is no counterpart to the
magnon-mediated repulsion of model I. Model III, on the other hand,
has only very weak interactions between its ``holes'', due to the fact
that if they are on neighbor sites they break fewer FM bonds then if
they are farther apart. This very weak attraction is of order $J$ and
describes completely different physics.

Since the magnon-mediated interactions that arise between carriers are
very different in the three models, it is clear that they do not
describe the same low-energy physics in cases with two or more
carriers, even though their quasiparticles are similar in nature. We
therefore conclude that the two-sublattice model cannot be replaced by
simple one-lattice models, whether with one or with more bands, because
these no longer allow two carriers to interact simultaneously with the
same lattice spin and, as a result, completely change the nature of the
magnon-mediated interactions.

\begin{figure}[t]
 \includegraphics[width=0.9\columnwidth]{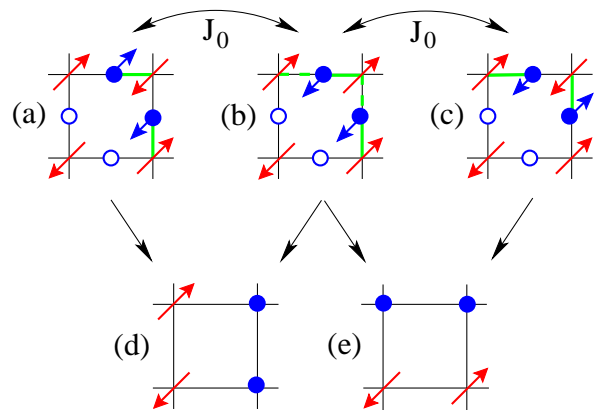}
  \caption{(Color online) (a), (b) and (c) depict two-hole
    configurations on a CuO$_2$ layer, connected to one another by
    spin-flip processes of order $J_0$. These are the 2D analogs of
    Fig. \ref{fig7}(a), and describe the main mechanism for 
    magnon-mediated attraction between holes. (d) and (e) show their
    corresponding one-band configurations, obtained by locking the
    holes in ZRS with a neighbor Cu spin. The thick light green
    lines in (a), (b) and (c) show which hole can pair with which
    spin. For (b), there are two possible pairing options, indicated
    by full and dashed thick green lines, respectively.  }
  \label{fig10}
\end{figure}

While each new problem has to be considered carefully, we expect that
this situation may be the rule rather
than an exception. The reason  is that the ZRS-like states on
which the simpler models are built mix together charge and spin
degrees of freedom. As such, it seems unlikely to us that the
magnon-mediated interactions between carriers could be easily
``mimicked'' by simple effective interactions between these complex
objects.

Even though all our exact results are for a FM background, we think that
simple arguments show that these claims extend to AFM
backgrounds. Consider, for instance the situation sketched in
Fig. \ref{fig10}(a), with two neighboring holes on a CuO$_2$-like
lattice. For simplicity, we assume that the Cu spins are
well-described by N\'eel order on the length scale of interest
here.

In the two-sublattice model, configuration (a) is connected through
matrix elements of order $J_0$ to configurations (b)
and (c). This is the 2D analog of the exchange of a magnon through the
common lattice spin depicted in Fig. \ref{fig7}(a), i.e. precisely the
process expected to be responsible for the magnon-mediated
interactions between carriers. If we  lock these holes into ZRS
and map configuration (a) to a one-band model, we obtain the state of
Fig. \ref{fig10}(d). Configuration (b) can map to the same one-band state
(d), because it is part of the same ZRS as (a), but it can also map to
configuration (e), depending on how we choose which hole pairs with
which spins. Finally, configuration
(c) maps to (e), as expected since it is part of the same ZRS as (b).

The problem is now obvious: the one-band $t$-$J$ model has no matrix
element between states (d) and (e), never mind one of order $J_0$ (this
energy scale has been removed from this model). As such, it clearly
fails to describe the same physics as the two-sublattice model, since
it grossly underestimates the strength of the magnon-mediated
attraction between carriers. Instead of this back-and-forth exchange
of the magnon that strongly favors configurations with neighbor
carriers, the $t$-$J$ model only describes weak interactions of order
$J$ between ``holes'' because of the number of AFM bonds broken as
they move around and reshuffle the lattice spins; this is completely
different physics. This example explains why we believe that simple
one-band models cannot be expected to describe properly interactions
between carriers.

Two possible ways to fix the problem can be envisioned. One is 
to add additional terms in the one-band Hamiltonians to generate the
missing matrix elements. For example, states (d) and (e) in
Fig. \ref{fig10} could be connected by a term describing
second-nearest neighbor hopping of a hole if there is a second hole in
its vicinity. Similar ``conditional'' third nearest-neighbor hopping
must also be included for configurations where the holes are on the
same line.  These terms would strongly favor pairing of the
carriers, especially since they do not disturb the AFM background
order. In fact, since $J_0$ is a large energy scale, this might 
explain the existence of rather mobile pre-formed pairs, which is one
of the favorite scenarios to explain the pseudo-gap regime. 

While this way to fix the one-band models is  appealing, care is
needed to make sure that all these additional matrix elements have the
proper signs (i.e., like those arising in the more complex model); and
also that the effects of other processes, such as the spin-swap
terms\cite{Bayo1} which we have not included in this model, are
properly accounted for and ``mimicked''. It should also be explored
what, if any, are the consequences of the fact that multiple
states in the one-band Hilbert space may correspond to a single
state of the two-sublattice model, as shown in
Fig. \ref{fig10} for configurations (b), (d) and (e) (note that there
is also a third one-band configuration corresponding to (b), with the
ZRS diagonally opposite to each other).

The alternative is to study the proper two-sublattice models. They
have larger Hilbert spaces, unfortunately, but at least they are
certain to
describe correctly the significant magnon-mediated attraction between
carriers, which may well be responsible for most, if not all, the glue
needed for superconductivity in cuprates.

\begin{acknowledgments} M. M. thanks Prof. P. Brouwer for facilitating
  the visiting appointment that lead to this collaboration, and the
PROMOS Scholarship of the FU Berlin. This work was supported by NSERC,
CIfAR and QMI.

\end{acknowledgments}

\appendix

\section{Details for the real-space solution for the two-carrier case}

The EOM for model I are:

\begin{widetext}
\begin{align}
 G(n,n',k,\omega)=& \frac{1}{\omega+i\eta} \left \{
  \vphantom{\sqrt{\frac{S}{2}}} \delta_{n,n'}-2t \cos \frac{k}{2}
  \left [ G(n+1,n',k,\omega)+G(n-1,n',k,\omega) \right]
  \right. &. \nonumber \\ & \left. +J_0 \sqrt{\frac{S}{2}} \left [
  G(n,n,n',k,\omega)+G(n,n+1,n',k,\omega)\right] \right \}, &
  \text{for} \ n>0 \label{eq:motionGn>0},\\ G(n,n',k,\omega)=&
  \frac{1}{\omega+i\eta} \left \{ \vphantom{\sqrt{\frac{S}{2}}}
  \delta_{n,n'}-2t \cos \frac{k}{2} \left [
  G(n+1,n',k,\omega)+G(n-1,n',k,\omega) \right] \right. &. \nonumber
  \\ & \left. -J_0 \sqrt{\frac{S}{2}} \left [
  G(-n,0,n',k,\omega)+G(-n,1,n',k,\omega)\right] \right \}, &
  \text{for} \ n<0 \label{eq:motionGn<0},\\ G(0,n',k,\omega)=&
  \frac{1}{\omega-U+i\eta} \left \{ \vphantom{\sqrt{\frac{S}{2}}}
  \delta_{0,n'}-2t \cos \frac{k}{2} \left [
  G(1,n',k,\omega)+G(-1,n',k,\omega) \right] \right \},
  \label{eq:motionGn=0}
\end{align}
\begin{align}
G(n,m,n',k,\omega)=&\frac{1}{\omega-2JS-\frac{J_0}{2}[4S-\delta_{m,0}-\delta_{m,1}-\delta_{m,n}-\delta_{m,n+1}]+i\eta}
& \nonumber \\ & \times \left \{ -t \left [ e^{-i\frac{k}{2}}
(1-\delta_{n,1})G(n-1,m-1,n',k,\omega) + e^{i\frac{k}{2}} G(n+1,m+1,n',k,\omega)
\right. \right. & \nonumber \\ &
\left. \left. \vphantom{e^{i\frac{k}{2}}} + e^{-i\frac{k}{2}}
G(n+1,m,n',k,\omega)+e^{i\frac{k}{2}} (1-\delta_{n,1})G(n-1,m,n',k,\omega) \right] -JS
\left [ G(n,m+1,n',k,\omega) \right. \right. & \nonumber \\ &
\left. \vphantom{e^{i\frac{k}{2}}} \left.  +
G(n,m-1,n',k,\omega)\right ] +J_0 \sqrt{\frac{S}{2}} [(\delta_{m,n}
+\delta_{m,n+1})G(n,n',k,\omega) \right. & \nonumber \\ &
\left. \vphantom{e^{\frac{k}{2}}} -(\delta_{m,0}+\delta_{m,1})
G(-n,n',k,\omega) ] \right \} \label{eq:motionGm}, \ \text{for}\ n
\geq 1,\ -\infty<m<\infty .
\end{align}
\end{widetext}

\section{Solution for $G(0,0,k,\omega)$}

We show here how to  calculate
$G(0,0,k,\omega)$. From Eq. (\ref{eq:motionGn=0}) we see that
$G(0,0,k,\omega)$ is linked only to $G(\pm 1,0,k,\omega)$. We can choose
to order the vectors $\mathbf{V}_{M}$ in such a way that
$G(1,0,k,\omega)=[\mathbf{A}_1]_{1,1} G(0,0,k,\omega)$ and
$G(-1,0,k,\omega)=[\mathbf{A}_1]_{2,1} G(0,0,k,\omega)$. Inserting
this into Eq. (\ref{eq:motionGn=0}) we find:
\begin{align}
%  &G(0,0,k,\omega)=\left [\omega+ i \eta -U +2 t \cos \frac{k}{2} ([\mathbf{A}_0]_{1,1}+[\mathbf{A}_0]_{2,1}) \right]^{-1} \\
 &G(0,0,k,\omega)=\frac{1}{\omega+ i \eta -U +2 t \cos \frac{k}{2}
 ([\mathbf{A}_1]_{1,1}+[\mathbf{A}_1]_{2,1}) }, \nonumber
\end{align}
where $\mathbf{A}_1$ is obtained from the recursive relation
Eq. (\ref{eq:recursive}). Once $G(0,0,k,\omega)$ and therefore
$\mathbf{V}_0$ is known, one can use the recursive relation
Eq. (\ref{eq:A}) to find  $\mathbf{V}_M=\mathbf{A}_M \dots
\mathbf{A}_1\mathbf{V}_0$, and thus  to generate
all Green's functions corresponding to values of $M>0$.

\section{Details for VS1 for model I, in the triplet-like sector}

We introduce the shorthand notation:
\begin{align}
 &a^+_{1}= _+\!\!\langle k, 1| \hat{G}(\omega)| k, 1\rangle_+,\\
%%%%%%%%%%%%%%%%%%%%%%%%%%%%%%%%%%%%%%%%%%%%%%%%%%%%%%%%%%%%%%%%%%%%%%%%%%%
 &a^+_{1,1}=_+\!\!\langle k, 1| \hat{G}(\omega)| k, 1,1\rangle_+,\\
%%%%%%%%%%%%%%%%%%%%%%%%%%%%%%%%%%%%%%%%%%%%%%%%%%%%%%%%%%%%%%%%%%%%%%%%%%%
 &a^+_{1,m}=_+\!\!\langle k, 1| \hat{G}(\omega)| k, 1,m\rangle_+ \
 \text{   for}\ m<1,
%%%%%%%%%%%%%%%%%%%%%%%%%%%%%%%%%%%%%%%%%%%%%%%%%%%%%%%%%%%%%%%%%%%%%%%%%%%
\intertext{and using Eqs (\ref{eq:motionGn>0})-(\ref{eq:motionGm}) we
obtain the following EOM:}
%%%%%%%%%%%%%%%%%%%%%%%%%%%%%%%%%%%%%%%%%%%%%%%%%%%%%%%%%%%%%%%%%%%%%%%%%%%
 &a^+_{1}=\frac{1+J_0\sqrt{\frac{S}{2}}(a^+_{1,0}+\sqrt{2}a^+_{1,1})}{\omega+i\eta},\\
%%%%%%%%%%%%%%%%%%%%%%%%%%%%%%%%%%%%%%%%%%%%%%%%%%%%%%%%%%%%%%%%%%%%%%%%%%%
 &a^+_{1,1}=\frac{J_0\sqrt{\frac{S}{2}}\sqrt{2}a^+_{1}-JS\sqrt{2}a^+_{1,0}}{\omega+i\eta-2JS-J_0(2S-1)},\\
%%%%%%%%%%%%%%%%%%%%%%%%%%%%%%%%%%%%%%%%%%%%%%%%%%%%%%%%%%%%%%%%%%%%%%%%%%%
 &a^+_{1,0}=\frac{J_0\sqrt{\frac{S}{2}}a^+_{1}-JS\sqrt{2}a^+_{1,1}-JSa^+_{1,-1}}{\omega+i\eta-2JS-J_0(2S-1/2)}.\label{eq:ElBetwSpinVariationalm}\\
%%%%%%%%%%%%%%%%%%%%%%%%%%%%%%%%%%%%%%%%%%%%%%%%%%%%%%%%%%%%%%%%%%%%%%%%%%%
 &
 a^+_{1,m}=\frac{-JSa^+_{1,m-1}-JSa^+_{1,m+1}}{\omega+i\eta-2JS-2J_0S},\
 \text{for}\ m<0. 
\end{align}
As appropriate for VS1, in writing the above equations we ignored all
Green's functions for basis states where 
the carrier-carrier distance $n\ge2$. 

The last of the above equations  is a simple recurrence equation that
can be solved with the ansatz 
$a^+_{1,m}(\omega)=z(\omega)a^+_{1,m+1}(\omega)$ (note that
$m<0$). This ansatz makes use of the fact that
$a_{m}^+(\omega)\xrightarrow {m\rightarrow -\infty}0$ and therefore
requires $|z(\omega)|<1$. As previously discussed this is true because
we are using a finite lifetime $1/\eta$. 

Then, using $a^\pm_{1,-1}(\omega)=z(\omega)a^\pm_{1,0}(\omega)$ in
Eq. (\ref{eq:ElBetwSpinVariationalm}) 
reduces the first 3 equations to a linear set with 3 unknowns,
$a^+_{1}, a^+_{1,1}$ and $a^+_{1,0}$. This can be solved
analytically. The solution (for $J=0$, so as to keep it compact) is:
\begin{align}
 a^+_1 &= \left \{\omega+i\eta -J_0^2 S \left [ \frac{1}{\omega+i\eta
 +J_0 (1-2S)} \right. \right. \nonumber \\ & \left. \left.
 +\frac{\frac{1}{2}}{\omega+i\eta +J_0 (\frac{1}{2}-2S)}\right ]
 \right \}^{-1}.
\end{align}

While an analytical solution is not possible in all cases, the
significant reduction in the number of unknowns allows us to find  very easily
numerical solutions for the various VS, and therefore
approximations for the energies of the bound states (if any exist).

\section{Effective mass of the bipolaron}

The first order correction is obtained by diagonalizing
$\mathcal{H}_{ex}^{(I)}$ in the subspace spanned by $| k, n=1\rangle +
| k, n=-1 \rangle, | k,
 n=1, m=1\rangle$ and $| k, n=1, m=0\rangle+ | k,n=1,
 m=2\rangle$.
Note that all of the above states have triplet-like symmetry and that
this subspace is invariant under $\mathcal{H}_{ex}^{(I)}$ since this
interaction cannot change the carrier-carrier distance. The lowest eigenvalue of
$\mathcal{H}_{ex}^{(I)}$ in this subspace gives the zeroth order
approximation for $E_{\text{T1}}$, while the first order correction in
$t$ and $J$ is given by the matrix elements of $\hat{T}$ and
$\mathcal{H}_{S}^{(I)}$ for the corresponding eigenvector. This
results in:
$$E_{\text{T1}}^{(0)}(k)=-\frac{J_0}{4} \left(3 -4S+\sqrt{1+16
 S^2}\right),
$$
$$
\Delta E_{\text{T1}}^{(1)}(k)=\frac{J S \left(2+\frac{5
 S \left(-1-4 S+\sqrt{1+16 S^2}\right)}{\sqrt{1+16 S^2}}\right)}{1+3
 S}.
$$
Note that (i) the first order correction in $t$ vanishes, since
$\hat{T}$ changes the carrier-carrier distance and therefore leaves
this subspace, 
and (ii) the zeroth order energy is identical to the VS1 result, see
Eq. (\ref{eq:ElBetwSpinVariationalGS}), as expected.

In order to obtain higher order corrections we need to include all the
states that are linked to by $\hat{T}$ and $\mathcal{H}_{S}^{(I)}$
when starting from the configurations considered above. If need be, we
also have to add whatever other configurations are necessary to ensure
that the enlarged subspace is still invariant under  
$\mathcal{H}_{ex}^{(I)}$. This means that we need to include a total
of seven additional states, some of which have singlet-like symmetry,
since at finite $k$ mixing between the two symmetries occurs. However,
for third order corrections one does not need to include the state
where the carriers are on the same site and therefore we do not need
to worry about $\hat{U}$. This may seem counterintuitive, since we are
starting out with states where the carriers are on adjacent sites,
{\em i.e.} $n=1$, and therefore one would expect a transition to
$n=0$ to be a simple process. However, the state with $n=0$ only
exists for singlet-like symmetry and since we start out with
triplet-like states, the transition $n=1\rightarrow n=0$ only
contributes corrections higher
than third order.

For the sake of simplicity we only write here the results obtained for
$S=\frac{1}{2}$: 
\begin{widetext}
\begin{align}
 &\Delta E_{\text{T1}}^{(2)}(k) = -\frac{\left(275+119 \sqrt{5}\right)
 J^2+40 t^2 \left(245+103 \sqrt{5}+\left(205+97 \sqrt{5}\right) \cos
 k\right)}{125 \left(3+\sqrt{5}\right)^2 J_0},\\
%%%%%%%%%%%%%%%%%%%%%%%%%%%%%%%%%%%%%%%%%%%%%%%%%%%%%%%%%%%%%%%%%%%%%%%%%%%%%%%%
 &\Delta E_{\text{T1}}^{(3)}(k) = \frac{\left(-91+125 \sqrt{5}\right)
 J^3+20 J t^2 \left(-166+20 \sqrt{5}+\left(-209+15 \sqrt{5}\right)
 \cos k\right)}{1250 J_0^2}.
\end{align}
Besides an overall shift, a dispersion proportional to $\cos(k)$
arises in the 2nd and 3rd order corrections. The effective bipolaron mass
can now be extracted easily.
\end{widetext}

\end{document}